\PassOptionsToPackage{english}{babel}
\documentclass[reprint,aps,pre,twocolumn,amssymb,amsfonts, superscriptaddress,longbibliography]{revtex4-1}  % for review and submission
\usepackage[english]{babel}
\usepackage{graphicx}  % needed for figures
\usepackage{dcolumn}   % needed for some tables
\usepackage{bm}        % for math
\usepackage{amssymb}   % for math
\usepackage{amsmath}
\usepackage{mathtools}
\usepackage{bbold}
\usepackage[utf8]{inputenc}
\usepackage{scalerel}
\usepackage[nokeyprefix]{refstyle}
\usepackage{varioref}
\usepackage{hyperref}
\usepackage{changes}
\usepackage[normalem]{ulem}

\definecolor{myurlcolor}{rgb}{0,0,0.7}
\definecolor{myrefcolor}{rgb}{0.8,0,0}
\usepackage{hyperref}
\hypersetup{colorlinks, linkcolor=myrefcolor,
citecolor=myurlcolor, urlcolor=myurlcolor}

\usepackage{cleveref}
\usepackage{comment}

\newcommand{\ie}{\textit{i.e. }}
\newcommand{\eg}{\textit{e.g. }}

\newcommand{\ignore}[1]{}

\newcommand{\dd}{\text{d}}

\begin{document}
\selectlanguage{english}

\title{Normalized Gaussian Path Integrals}

\author{Giulio Corazza}\email{giulio.corazza@epfl.ch} \affiliation{Laboratory for Computation and Visualization in Mathematics and Mechanics (LCVMM) Institute of Mathematics, Swiss Federal Institute of Technology (EPFL), CH-1015 Lausanne, Switzerland} 
\author{Matteo Fadel}\email{matteo.fadel@unibas.ch} \affiliation{Department of Physics,
University of Basel, Klingelbergstrasse 82, 4056 Basel, Switzerland} 

\date{\today}

\begin{abstract}
Path integrals play a crucial role in describing the dynamics of physical systems subject to classical or quantum noise. In fact, when correctly normalized, they express the probability of transition between two states of the system. In this work, we show a consistent approach to solve conditional and unconditional Euclidean (Wiener) Gaussian path integrals that allow us to compute transition probabilities in the semi-classical approximation from the solutions of a system of linear differential equations. 
Our method is particularly useful for investigating Fokker-Planck dynamics, and the physics of string-like objects such as polymers.
To give some examples, we derive the time evolution of the $d$-dimensional Ornstein-Uhlenbeck process, and of the Van der Pol oscillator driven by white noise. Moreover, we compute the end-to-end transition probability for a charged string at thermal equilibrium, when an external field is applied.
\end{abstract}

\maketitle

\section{Introduction}

Path integrals are an essential tool in many branches of physics and mathematics \cite{BookPapa,BookSchulman,BookChaichian,BookZinn,BookKleinert}. Originally introduced by Wiener as a method to study Brownian motion \cite{Wiener1,Wiener2}, their formalism was significantly developed by Feynman in the context of quantum mechanics \cite{FEY0,FEY}. Since then, path integrals revealed themselves to be a powerful method for the investigation of systems subject to classical or quantum fluctuations, therefore finding a plethora of different applications.\\
\indent In many relevant situations one is interested in evaluating transition (\ie conditional) probabilities. Namely, the probability for a system to be in a specific final state, given its initial state. Path integrals are precisely tailored to answer such questions, by expressing transition probabilities as an infinite weighted sum over all possible trajectories passing through both states. Typical examples where this formulation arises naturally include the stochastic motion of particles in diffusion processes, and the dynamics of quantum particles and fields.\\
\indent It is worth emphasizing that the ``paths'' entering a path integral do not need to be the trajectories of a moving particle, but they can also be the stationary configurations of string-like objects \cite{Edwards1965,Edwards1967,Freed,PapaPoly}. In this context, transition probabilities represent the probability of finding the string's endpoints in specific positions. This observation turns out extremely useful for the study of organic and inorganic polymers at thermal equilibrium, such as chains of molecules (\eg DNA, actin filaments) and flexible rods \cite{Winkler94,Winkler97,Vilgis00,LUD}.\\
\indent Despite their intuitive interpretation, path integrals are in general difficult to compute. Among several different strategies to circumvent this issue, the semiclassical (quadratic) approximation is one of the most adopted \cite{BookLan,REC,MOR}. In brief, the idea consists in approximating the weights for the paths so that a Gaussian integral is obtained. The solution is then straightforward for conditional path integrals (where both extremal points are fixed), while it often remains non-trivial for the unconditional case (where only the starting point is fixed). Addressing this remaining problem is of special interest for expressing transition probability distributions that are properly normalized.\\
\indent Here we focus on Euclidean (Wiener) path integrals, and propose a consistent method to compute from them transition probabilities in the semiclassical approximation. Our approach is based on the generalization of a result by Papadopoulos \cite{PAP1}, which allows us to evaluate both conditional and unconditional path integrals for general quadratic Lagrangians, from the solutions of the Euler-Lagrange equations and of a system of second-order nonlinear differential equations. Furthermore, we then show that the latter can be related to a simpler system of linear differential equations, by exploiting a link with the Jacobi equation.
Interestingly, our study also sheds light on the relation between the choice for the discretization of continuous paths, and the path integral measure.\\
\indent Our results are of interest for studying the dynamics of stochastic processes, such as the one described by the Fokker-Planck equation, and for investigating equilibrium configurations of string-like objects. This is illustrated here with three concrete examples.
First, we show how to recover the transition probability for a $d$-dimensional Ornstein-Uhlenbeck process \cite{OU,FALKOFF,VATI}. Second, we investigate the non-linear Van der Pol oscillator driven by white noise, for which transition probabilities are not known analytically due to its chaotic dynamics \cite{NAESS}. Third, we compute in one spatial dimension the end-to-end transition probability for an elastic and electrically charged string at thermal equilibrium, when an external electric field is applied.

\section{Statement of the problem and main results}

Consider a system in configuration $q(\tau)\in\mathbb{R}^{d}$, whose dynamics is described by the Lagrangian $\mathcal{L}(q,\dot{q},\tau)$. Our goal is to calculate the transition probability for the system of being in final state $q(t)=q$, given its initial state $q(t_0)=q_0$, namely the conditional probability $\rho(q,t \vert q_0,t_0)$ satisfying $\rho(q,t_0 \vert q_0,t_0)=\delta(q-q_0)$. A prescription for this calculation is given by the path integral formalism, which allows us to write
\begin{equation}\label{pathint}
\rho(q,t \vert q_0,t_0) := \dfrac{\mathcal{K}}{\mathcal{N}} = \frac{1}{\mathcal{N}}\int\limits_{q(t_0)=q_0}^{q(t)=q}{\mathcal{D}q\,e^{-S(q,\dot{q})}} \;,
\end{equation}%
where the (conditional) integration is taken over all paths with fixed extremal points, weighted depending on the action $S(q,\dot{q}):=\int_{t_0}^t{d\tau\,\mathcal{L}(q,\dot{q},\tau)}$, and normalized by $\mathcal{N}$ to ensure
\begin{equation}\label{norm}
\int_{\mathbb{R}^{d}}{\rho(q,t \vert q_0,t_0)}\,\dd q = 1 \;, \qquad \forall t\geq t_0 \;.
\end{equation}
From Eqs.~(\ref{pathint}) and (\ref{norm}), we see that the normalization can be formally written as the (unconditional) path integral 
\begin{equation}\label{normex}
\mathcal{N} = \int\limits_{q(t_0)=q_0}{\mathcal{D}q\,e^{-S(q,\dot{q})} } \;,
\end{equation}%
where now the integral is over all paths satisfying only the initial condition $q(t_0)=q_0$.

For typical cases of interest, we are often in the situation where $\mathcal{L}$ is complicated enough that closed-form solutions for $\mathcal{K}$ and $\mathcal{N}$ do not exist. A standard technique to simplify part of the problem consist in taking the semi-classical approximation, where the action is expanded to second order around an isolated minimum. This allows us to approximate $\mathcal{K}$ by a solvable Gaussian integral, but the evaluation of $\mathcal{N}$ remains non-trivial because of the free boundary condition $q(t)$. The latter difficulty is often circumvented through demanding Monte-Carlo integrations, or by setting $\mathcal{N}=1$ and considering in $\mathcal{K}$ an effective (Onsager-Machlup) Lagrangian containing additional terms that ensure normalization \cite{GRAHAM, FALKOFF, HAKEN}. Our main result consists in solving this problem in a more general situation. In brief, we formulate a consistent approach to solve in the semi-classical approximations both path integrals appearing in Eq.~(\ref{pathint}), by relating their solutions to the solutions of a system of linear differential equations. The procedure we propose is the following.

As a starting point, in order to ensure the accuracy of the semi-classical approximation, let us restrict to Lagrangian functions where the leading order term for the second variation of the action, $G(\tau):= \frac{\partial^2 \mathcal{L}}{\partial \dot{q}^2}$, is independent of $q, \dot{q}$. This assumption is still general enough to include most cases of interest. On the other hand, we consider in the second variation arbitrary $V:= \frac{\partial^2 \mathcal{L}}{\partial {q}^2}$ and cross term matrix $A:= \frac{\partial^2 \mathcal{L}}{\partial \dot{q}\partial q}$ not necessarily symmetric.

Following the idea behind the semi-classical approximation, the first step of our method consists in deriving from the Euler-Lagrange equations for $\mathcal{L}$ two solutions: 
\begin{enumerate}
\item an isolated minimizer of the action $q^{\scaleto{D}{3.5pt}}(\tau)$, satisfying the Dirichlet boundary conditions $q^{\scaleto{D}{3.5pt}}(t_0)=q_0$ and $q^{\scaleto{D}{3.5pt}}(t)=q$,
\item an isolated minimizer $q^{\scaleto{N}{3.5pt}}(\tau)$, satisfying $q^{\scaleto{N}{3.5pt}}(t_0)=q_0$ and the Neumann natural boundary condition $\frac{\partial \mathcal{L}}{\partial \dot{q}}(t)=0$.
\end{enumerate}
Then, the second step of our method consists in deriving a set of solutions $W^{\scaleto{D}{3.5pt}(\scaleto{N}{3.5pt})}(\tau)$ of the Jacobi equation for the second variation of the action on the Dirichlet (Neumann) minimum. These can be obtained from the Hamiltonian formulation of the Jacobi equation, together with the appropriate boundary conditions, as solutions of 
\begin{small}
\begin{equation}\label{fin3}
\begin{cases}
             \frac{\dd}{\dd\tau}\begin{pmatrix}
      {{W^{\scaleto{D}{3.5pt}}}}  \\
     {{M^{\scaleto{D}{3.5pt}}}}
     \end{pmatrix}
       =J{{E^{\scaleto{D}{3.5pt}}}}\begin{pmatrix}
       {{W^{\scaleto{D}{3.5pt}}}}  \\
     {{M^{\scaleto{D}{3.5pt}}}}
     \end{pmatrix}
   \\
   \\
\begin{pmatrix}
      {{W^{\scaleto{D}{3.5pt}}}}  \\
     {{M^{\scaleto{D}{3.5pt}}}}
     \end{pmatrix}
     (t)
       =J\begin{pmatrix}
       {\mathbb{1}} \\
     \mathbb{0}
     \end{pmatrix}
       \end{cases} \,
\begin{cases}
           \frac{\dd}{\dd\tau}\begin{pmatrix}
      {{W^{\scaleto{N}{3.5pt}}}}  \\
     {{M^{\scaleto{N}{3.5pt}}}}
     \end{pmatrix}
       =J{{E^{\scaleto{N}{3.5pt}}}}\begin{pmatrix}
       {{W^{\scaleto{N}{3.5pt}}}}  \\
     {{M^{\scaleto{N}{3.5pt}}}}
     \end{pmatrix}
   \\
   \\
\begin{pmatrix}
      {{W^{\scaleto{N}{3.5pt}}}}  \\
     {{M^{\scaleto{N}{3.5pt}}}}
     \end{pmatrix}
     (t)
       =J\begin{pmatrix}
     \mathbb{0} \\
     {\mathbb{1}} 
     \end{pmatrix}
       \end{cases}
\end{equation}
\end{small}%
where $M^{\scaleto{D}{3.5pt}(\scaleto{N}{3.5pt})}$ is the conjugate variable under the Legendre transform, $J=\bigl( \begin{smallmatrix} \mathbb{0} & \mathbb{1}\\ -\mathbb{1} & \mathbb{0}\end{smallmatrix}\bigr)$ is the symplectic matrix, and $E^{\scaleto{D}{3.5pt}(\scaleto{N}{3.5pt})}$ is the symmetric matrix driving the system, which reads
\begin{equation}\label{driv}
E^{\scaleto{i}{5pt}} = \begin{pmatrix}
     {A}^T{G}^{-1}{A}-{V}\,\,\, & -{A}^T{G}^{-1}\\
       -{G}^{-1}{A}\,\,\, & {G}^{-1} 
       \end{pmatrix}\Bigg|_{q^{\scaleto{i}{5pt}}}\;, \quad {\scaleto{i}{5pt}}={\scaleto{D}{3.5pt},\scaleto{N}{3.5pt}} \;.
\end{equation}

Finally, our first main result consists in showing that we can write the semi-classical approximation of the transition probability Eq.~(\ref{pathint}) as
\begin{equation}\label{fin2}
\rho^{\text{sc}}(q,t \vert q_0,t_0) =  e^{S(q^{\scaleto{N}{3pt}})-S(q^{\scaleto{D}{3pt}})}\sqrt{\det\left[\frac{1}{2\pi}\frac{{W^{\scaleto{N}{3.5pt}}}}{{W^{\scaleto{D}{3.5pt}}}}(t_0)\right]} \;,
\end{equation}
where $S(q^{\scaleto{D}{3.5pt}(\scaleto{N}{3.5pt})})$ is the action evaluated on the Dirichlet (Neumann) minimum, and $W^{\scaleto{D}{3.5pt}(\scaleto{N}{3.5pt})}$ are the solutions of Eq.~(\ref{fin3}).

In general, due to the semi-classical approximation, we have $\rho(q,t \vert q_0,t_0) \approx \rho^{\text{sc}}(q,t \vert q_0,t_0)$. 
However, let us mention that in the particular case where the Lagrangian is a quadratic function of $q$ and $\dot{q}$, then no error is introduced by the semi-classical approximation, and $\rho(q,t \vert q_0,t_0) = \rho^{\text{sc}}(q,t \vert q_0,t_0)$.
In the latter case the matrix $E$ is now independent of the particular minimum, and Eq.~(\ref{fin3}) simplifies further to
\begin{small}
\begin{equation}\label{fin5}
\begin{cases}
             \frac{d}{d\tau}\begin{pmatrix}
      {{W^{\scaleto{D}{3.5pt}}}} &  {{W^{\scaleto{N}{3.5pt}}}}\\
     {{M^{\scaleto{D}{3.5pt}}}}  &  {{M^{\scaleto{N}{3.5pt}}}}
     \end{pmatrix}
       =J{{E}}\begin{pmatrix}
       {{W^{\scaleto{D}{3.5pt}}}} &  {{W^{\scaleto{N}{3.5pt}}}}\\
     {{M^{\scaleto{D}{3.5pt}}}}  &  {{M^{\scaleto{N}{3.5pt}}}}
     \end{pmatrix}
   \\
   \\
\begin{pmatrix}
    {{W^{\scaleto{D}{3.5pt}}}} &  {{W^{\scaleto{N}{3.5pt}}}}\\
     {{M^{\scaleto{D}{3.5pt}}}}  &  {{M^{\scaleto{N}{3.5pt}}}}
     \end{pmatrix}
     (t)
       =J
       \end{cases} \;.
\end{equation}
\end{small}%

In addition, as a second main result, we present a generalization of Eq.~(\ref{fin2}) that allows us to compute marginal transition probabilities defined as it follows. We reorder the configuration variables as $q(\tau)=(q_{_V}(\tau),q_{_F}(\tau))\in\mathbb{R}^{d}$, where $q_{_V}(\tau):=(q_1,...,q_l)(\tau)\in\mathbb{R}^l$ and $q_{_F}(\tau):=(q_{l+1},...,q_d)(\tau)\in\mathbb{R}^{d-l}$, and consider the marginals $\rho_{\scaleto{\mathcal{M}}{3.5pt}}(q_{_F},t \vert q_0,t_0) := \int \text{d}q_{_V} \rho(q,t \vert q_0,t_0)$. The latter can be expressed as the path integral
\begin{equation}\label{pathint2}
\rho_{\scaleto{\mathcal{M}}{3.5pt}}(q_{_F},t \vert q_0,t_0) := \dfrac{\mathcal{K}_{\scaleto{\mathcal{M}}{3.5pt}}}{\mathcal{N}} = \frac{1}{\mathcal{N}}\int\limits_{q(t_0)=q_0}^{q_{_F}(t)=q_{_F}}{\mathcal{D}q\,e^{-S(q,\dot{q})}} \;,
\end{equation}%
where the integration for $K_{\scaleto{\mathcal{M}}{3.5pt}}$ is taken over all paths starting at $q(t_0)=q_0$ and with the mixed end-point conditions $q_{_F}(t)=q_{_F}$ fixed and $q_{_V}(t)$ variable. Note that the normalization term $\mathcal{N}$ remains the same as in Eq.~(\ref{normex}).

To derive a generalization of Eq.~(\ref{fin2}) for Eq.~(\ref{pathint2}) we follow the same strategy as before, and start by computing an isolated minimizer $q^{\scaleto{DN}{3.5pt}}(\tau)$ that satisfies the Euler-Lagrange equations with $q^{\scaleto{DN}{3.5pt}}(t_0)=q_0$, $q_{_F}^{\scaleto{DN}{3.5pt}}(t)=q_{_F}$ and with Neumann natural boundary condition for the remaining variables $\frac{\partial \mathcal{L}}{\partial \dot{q}_V}(t)=0$. Then, the semi-classical approximation of the transition probability Eq.~(\ref{pathint2}) is 
\begin{equation}\label{fin22}
\rho_{\scaleto{\mathcal{M}}{3.5pt}}^{\text{sc}}(q_{_F},t \vert q_0,t_0) =  e^{S(q^{\scaleto{N}{3pt}})-S(q^{\scaleto{DN}{3pt}})}\sqrt{(2\pi)^{l-d}\det\left[\frac{{W^{\scaleto{N}{3.5pt}}}}{{W^{\scaleto{DN}{3.5pt}}}}(t_0)\right]} \;,
\end{equation}
with $W^{\scaleto{DN}{3.5pt}}$ obtained from
\begin{small}
\begin{equation}\label{fin33}
\begin{cases}
           \frac{\dd}{\dd\tau}\begin{pmatrix}
      {{W^{\scaleto{DN}{3.5pt}}}}  \\
     {{M^{\scaleto{DN}{3.5pt}}}}
     \end{pmatrix}
       =J{{E^{\scaleto{DN}{3.5pt}}}}\begin{pmatrix}
       {{W^{\scaleto{DN}{3.5pt}}}}  \\
     {{M^{\scaleto{DN}{3.5pt}}}}
     \end{pmatrix}
   \\
   \\
\begin{pmatrix}
      {{W^{\scaleto{DN}{3.5pt}}}}  \\
     {{M^{\scaleto{DN}{3.5pt}}}}
     \end{pmatrix}
     (t)
       =\begin{pmatrix}
     \mathbb{1}_{l\times l} & \mathbb{0}_{d+l\times d-l}\\
      \mathbb{0}_{2d-l\times l} & -\mathbb{1}_{d-l\times d-l}
       \end{pmatrix}
       \end{cases} \;,
\end{equation}
\end{small}\\
where $M^{\scaleto{DN}{3.5pt}}$ is the conjugate variable under the Legendre transform for the Jacobi equation in Hamiltonian form, and $E^{\scaleto{DN}{3.5pt}}$ is the symmetric matrix Eq.~(\ref{driv}) here computed on the mixed minimum $q^{\scaleto{DN}{3.5pt}}(\tau)$. Note that for $l=0$ we have that Eq.~(\ref{fin22}) reduces to Eq.~(\ref{fin2}), since $q^{\scaleto{DN}{3.5pt}}(\tau)$ becomes $q^{\scaleto{D}{3.5pt}}(\tau)$ for the boundary conditions of the Euler-Lagrange equations, and Eq.~(\ref{fin33}) reduces to the the first system for $W^{\scaleto{D}{3.5pt}}$ in Eq.~(\ref{fin3}). On the other hand, for $l=d$ we have that $\mathcal{K}_{\scaleto{\mathcal{M}}{3.5pt}}$ coincides with the normalization factor $\mathcal{N}$.

To summarize, our approach for computing the transition probabilities Eq.~(\ref{pathint}) and Eq.~(\ref{pathint2}) consists in taking the ratio of the conditional and unconditional Wiener path integrals in the semi-classical approximation, to then express their solutions in terms of the solutions of a set of ordinary differential equations. In the following we present the derivation of our results, and we apply them to three relevant examples.
\vspace{3mm}

\section{Description of the method}

The evaluation of $\mathcal{K}$ in the semi-classical approximation is a standard textbook technique, and for our Euclidean path integrals it is also known as Laplace asymptotic method \cite{PIT}. The idea is to first Taylor expand the action $S(q(\tau))$ to second order around the Dirichlet minimum $q^{\scaleto{D}{3.5pt}}(\tau)$, exploiting the fact that the first order variation on a minimum is zero. Here, the existence and stability of $q^{\scaleto{D}{3.5pt}}(\tau)$ are assumed. In particular, the second property involves \eg the conjugate point theory, as discussed in \cite{FOM}. Then, from the second variation of the action computed in $q^{\scaleto{D}{3.5pt}}(\tau)$, namely
\begin{small}
\begin{equation}\label{secv}
\delta^2 S(q^{\scaleto{D}{3pt}},y) = \int_{t_0}^t \dd\tau\, \bigg( \dot{y}^T{{G}}\dot{y}+2\dot{y}^T{{A^{\scaleto{D}{3.5pt}}}}{y}+{y}^T{{V^{\scaleto{D}{3.5pt}}}}{y}\bigg) \;,
\end{equation}
\end{small}%
and from the discretisation of $\tau\in[t_0,t]$ into $n$ intervals of length $\varepsilon=(t-t_0)/n$, the semi-classical approximation for $\mathcal{K}$ reads
\begin{equation}\label{appx}
\mathcal{K}^{\text{sc}} = e^{-S(q^{\scaleto{D}{3pt}})} \lim_{n\rightarrow\infty} I^{\scaleto{D}{3.5pt}}_n \;,
\end{equation}
\begin{small}
\begin{equation}\label{num}
I^{\scaleto{D}{3.5pt}}_n=\displaystyle{\int\limits_{y(t_0)=0}^{y(t)=0}{\prod\limits_{j=1}^{n}\left[\frac{\det{(G_j)}}{(2\pi\varepsilon)^{d}}\right]^{\frac{1}{2}}\prod\limits_{j=1}^{n-1}{\dd y_j}\,\,\, e^{-\frac{\varepsilon}{2} \sum\limits_{j=0}^{n}\delta^2 S(q^{\scaleto{D}{3pt}},y)_j}}}.
\end{equation}
\end{small}%
The subscript $j$ indicates that the associated term is evaluated in $\tau_j=t_0+j\varepsilon$, \eg $y_j=y(\tau_j)$. Moreover, the integration boundaries come from the fact that $y(\tau)$ represents a perturbation around the minimum $q^{\scaleto{D}{3.5pt}}(\tau)$, and as such it must satisfy null Dirichlet boundary conditions. Let us mention that the products in Eq.~(\ref{num}) give the integration measure for the integral, which is here a conditional Wiener measure \cite{BookChaichian}. 

At this point, it is straightforward to solve Eq.~(\ref{appx}) using the method presented by Papadopoulos in Ref.~\cite{PAP1}. This results in the following Gelfand–Yaglom-type expression 
\begin{equation}\label{pap}
\mathcal{K}^{\text{sc}} = e^{-S(q^{\scaleto{D}{3pt}})}\det\left[2\pi {D^{\scaleto{D}{3.5pt}}}(t)\right]^{-\frac{1}{2}} \;,
\end{equation}
where ${D^{\scaleto{D}{3.5pt}}}(\tau)$ solves the second order nonlinear differential equation (omitting superscripts $\scaleto{D}{4.5pt}$ for $A$, $V$ and $D$)
\begin{small}
\begin{equation}\label{papeq}
\begin{split}
\frac{d}{d\tau}\left[\dot{D}{{G}}\right]+{{D}}\dot{A}^{(s)}-{{D}}\left[{{V}}+{{A}}^{(a)}{{G}}^{-1}{{A}}^{(a)}\right]=\\
\dot{D}{{A}}^{(a)}-{{D}}{{A}}^{(a)}{{G}}^{-1}{{D}}^{-1}\dot{D}{{G}} \;,
\end{split}
\end{equation}
\end{small}%
with ${A}^{(s)}$ $({A}^{(a)})$ the (anti-)symmetric part of $A^{\scaleto{D}{3.5pt}}$, and with initial conditions ${D^{\scaleto{D}{3.5pt}}}(t_0)=\mathbb{0}$, ${\dot{D}^{\scaleto{D}{3.5pt}}}(t_0)=G(t_0)^{-1}$.

We point out that the result Eq.~(\ref{pap}) is specific to the (Stratonovich-type) discretisation prescription adopted in Ref.~\cite{PAP1} for the cross terms $2\dot{y}(\tau)^T{{A}}(\tau){y}(\tau)$, which gives $\frac{1}{\varepsilon}(y_{j+1}-y_j)^T(A_jy_j+A_{j+1}y_{j+1})$. In fact, there is in general a one-parameter family of discretisations
\begin{equation}\label{disc}
\frac{2}{\varepsilon}(y_{j+1}-y_j)^T((1-\gamma)A_jy_j+\gamma A_{j+1}y_{j+1}),\,\,\,\gamma\in[0,1]
\end{equation}
leading to different results for Eq.~(\ref{pap}), \cite{LANG1,LANG2}. Interestingly, we notice that the mid-point rule ($\gamma=1/2$) is the only one giving a finite result for Eq.~(\ref{num}) when $A(\tau)$ is not symmetric. For more details see Appendix \ref{appB}.

It is now easy to show that, even for simple quadratic Lagrangians, Eq.~(\ref{pap}) alone does not represent a transition probability satisfying the normalization condition Eq.~(\ref{norm}). This can happen even if there are no cross terms (\ie $A=\mathbb{0}$), as we will see in the string example. To fix this issue, the condition Eq.~(\ref{norm}) is enforced by introducing the normalization factor $\mathcal{N}$, see Eq.~(\ref{pathint}). Unfortunately, computing $\mathcal{N}$ can be a non-trivial task, which we are now going to tackle.

Following the same approach as for $\mathcal{K}$, we compute the semi-classical approximation for $\mathcal{N}$ as defined in Eq.~(\ref{normex}). This time we Taylor expand the action $S(q(\tau))$ to second order around the Neumann minimum $q^{\scaleto{N}{3.5pt}}(\tau)$, since the point $q(t)$ is unconstrained. Then, from the second variation of the action computed in $q^{\scaleto{N}{3.5pt}}(\tau)$, namely
\begin{small}
\begin{equation}\label{secv2}
\delta^2 S(q^{\scaleto{N}{3pt}},h) = \int_{t_0}^t \dd\tau\, \bigg( \dot{h}^T{{G}}\dot{h}+2\dot{h}^T{{A^{\scaleto{N}{3.5pt}}}}{h}+{h}^T{{V^{\scaleto{N}{3.5pt}}}}{h}\bigg) \;,
\end{equation}
\end{small}%
and the same discretisation as before, the semi-classical approximation for $\mathcal{N}$ reads
\begin{equation}\label{appx2}
\mathcal{N}^{\textit{sc}} = e^{-S(q^{\scaleto{N}{3pt}})} \lim_{n\rightarrow\infty} I^{\scaleto{N}{3.5pt}}_n \;,
\end{equation}
\begin{small}
\begin{equation}\label{den}
I^{\scaleto{N}{3.5pt}}_n=\displaystyle{\int\limits_{h(t_0)=0}{\prod\limits_{j=1}^{n}\left[\frac{\det{(G_j)}}{(2\pi\varepsilon)^{d}}\right]^{\frac{1}{2}}\prod\limits_{j=1}^{n}{\dd h_j}\,\,\, e^{-\frac{\varepsilon}{2} \sum\limits_{j=0}^{n}\delta^2 S(q^{\scaleto{N}{3pt}},h)_j}}}.
\end{equation}
\end{small}%
Here, similarly to Eq.~(\ref{num}), the subscript $j$ indicates that the associated term is evaluated in $\tau_j$, and the integration boundaries come from the fact that $h(\tau)$ represents a perturbation around the minimum $q^{\scaleto{N}{3.5pt}}(\tau)$, and as such it must satisfy only the initial null Dirichlet boundary condition. The products in Eq.~(\ref{den}) give the integration measure for the integral, which is here an unconditional Wiener measure \cite{BookChaichian}. In fact, note that contrary to Eq.~(\ref{num}) the product of $\dd h_j$ runs here until $n$, which is what makes the integration of Eq.~(\ref{den}) in general nontrivial. 

One of our main results is now to show how to compute Eq.~(\ref{den}), from a modification of the method used by Papadopoulos in Ref.~\cite{PAP1} to derive Eq.~(\ref{pap}). The idea consists in performing a backward integration of Eq.~(\ref{den}), meaning that the standard direction of discretisation $(q_0,t_0)\rightarrow (q,t)$ is now replaced by $(q,t)\rightarrow (q_0,t_0)$. In summary, we are able to find a set of symmetric and positive definite matrices $F^{\scaleto{N}{3.5pt}}_j$, depending on $G$, $A^{\scaleto{N}{3.5pt}}$ and $V^{\scaleto{N}{3.5pt}}$, such that (see Appendix~\ref{Ap1intN})
\begin{equation}\label{res1}
{I}^{\scaleto{N}{3.5pt}}_n=\det{\left[\prod\limits_{j=1}^{n} F^{\scaleto{N}{3.5pt}}_j\right]}^{-\frac{1}{2}} \;. 
\end{equation}
Using then the recursion relations for ${I}^{\scaleto{N}{3.5pt}}_n$ and $F^{\scaleto{N}{3.5pt}}_j$, we show that the limit in Eq.~(\ref{res1}) gives
\begin{equation}\label{pap2}
\mathcal{N}^{sc} = e^{-S(q^{\scaleto{N}{3pt}})}\det\left[{D^{\scaleto{N}{3.5pt}}}(t_0)\right]^{-\frac{1}{2}}  \;,
\end{equation}
where $D^{\scaleto{N}{3.5pt}}(\tau)$ solves Eq.~(\ref{papeq}), but this time with $A^{\scaleto{N}{3.5pt}}(\tau)$ and $V^{\scaleto{N}{3.5pt}}(\tau)$. As a consequence of the backward integration necessary for deriving the matrices $F^{\scaleto{N}{3.5pt}}_j$, also Eq.~(\ref{papeq}) is now solved in the backward direction with boundary conditions ${D^{\scaleto{N}{3.5pt}}}(t)=\mathbb{1}$, ${\dot{D}^{\scaleto{N}{3.5pt}}}(t)=-{A^{\scaleto{N}{3.5pt}}}^{\small{(s)}}(t)G(t)^{-1}$.
Detailed calculations to derive this result are given in Appendix~\ref{Ap1limitN}.

Inspired by this strategy we compute again Eq.~(\ref{appx}), but this time using the backward integration procedure. We find that the result coincides with Eq.~(\ref{pap}), where now $D^{\scaleto{D}{3.5pt}}(\tau)$ solves Eq.~(\ref{papeq}) in the backward direction, with boundary conditions ${D^{\scaleto{D}{3.5pt}}}(t)=\mathbb{0}$, ${\dot{D}^{\scaleto{D}{3.5pt}}}(t)=-G(t)^{-1}$.

In conclusion, the results obtained so far allow us to express the transition probability Eq.~(\ref{pathint}) in the semi-classical approximation as
\begin{equation}\label{fin1}
\rho^{\text{sc}}(q,t \vert q_0,t_0):=\frac{\mathcal{K}^{\text{sc}}}{\mathcal{N}^{\text{sc}}} = e^{S(q^{\scaleto{N}{3pt}})-S(q^{\scaleto{D}{3pt}})}\sqrt{\det\left[\frac{1}{2\pi}\frac{{D^{\scaleto{N}{3.5pt}}}}{{D^{\scaleto{D}{3.5pt}}}}(t_0)\right]} \;.
\end{equation}

Despite the simplicity of Eq.~(\ref{fin1}), let us remember that the $D$'s have to be found by solving two second order nonlinear differential equations of the form of Eq.~(\ref{papeq}), which can be a demanding task. Remarkably, we are able to simplify this problem significantly by relating Eq.~(\ref{papeq}) to a system of linear differential equations.

As noted in Ref.~\cite{LUD}, it turns out that there is a relation between the matrix Eq.~(\ref{papeq}) and the linear Jacobi equation for a vector field ${w}\in\mathbb{R}^d$
\begin{equation}\label{jac}
\frac{d}{d\tau}\left[{G}\dot{w}+{A}{w}\right]-{A}^T \dot{w}-{V}{w}={{0}} \;.
\end{equation}
In fact, if $W=L^T$ is a matrix whose columns $w$ are solutions of the Jacobi equation Eq.~(\ref{jac}), then the solutions of Eq.~(\ref{papeq}) and the ones of
\begin{equation}\label{jactr}
\frac{d}{d\tau}\left[\dot{L}{G}+{L}{A}^T\right]-\dot{L}{A}-{L}{V}={\mathbb{0}}
\end{equation}
are related by the nonlinear transformation
\begin{equation}\label{tran}
L^{-1}\dot{L}=D^{-1}\dot{D}+{{A}}^{(a)}G^{-1} \;.
\end{equation}

Then, it is possible to impose the condition $\det{(L)}=\det{(D)}$, which is also equal to $\det{(W)}$, to ensure the uniqueness of the change of variables and to find the associated boundary conditions for the Jacobi equation. These read ${W^{\scaleto{D}{3.5pt}}}(t)=\mathbb{0}$, $ {\dot{W}^{\scaleto{D}{3.5pt}}}(t)=-G(t)^{-1}$ for $\mathcal{K}^{\text{sc}}$, and 
${W^{\scaleto{N}{3.5pt}}}(t)=\mathbb{1}$, ${\dot{W}^{\scaleto{N}{3.5pt}}}(t)=-G(t)^{-1}{A}(t)$ for $\mathcal{N}^{\text{sc}}$.  More details on this transformation are given in Appendix~\ref{appC}.

As a final step, since the Jacobi equation is the Euler-Lagrange equation for the second variation of the action, we can easily provide the more elegant Hamiltonian formulation given in Eq.~(\ref{fin3}). Namely, if $m=G\dot{w}+Aw$ is the conjugate variable under the Legendre transform of the second variation with respect to $\dot{w}$, and if we define $M=G\dot{W}+AW$, we can express Eq.~(\ref{fin1}) in terms of the solutions $W^{\scaleto{D}{3.5pt}}$ and $W^{\scaleto{N}{3.5pt}}$ of the Jacobi equation in Hamiltonian form subjected to the transformed final boundary conditions, \ie Eq.~(\ref{fin3}). This results then in Eq.~(\ref{fin2}).

To extend this result to marginal distributions Eq.~(\ref{pathint2}) we first explain how to combine the techniques involved in the computation of $\mathcal{K}$ and $\mathcal{N}$ in order to evaluate the path integral defining $\mathcal{K}_{\scaleto{\mathcal{M}}{3.5pt}}$. The semi-classical approximation for $\mathcal{K}_{\scaleto{\mathcal{M}}{3.5pt}}$ reads
\begin{equation}\label{appxDN}
\mathcal{K}_{\scaleto{\mathcal{M}}{3.5pt}}^{\text{sc}} = e^{-S(q^{\scaleto{DN}{3pt}})} \lim_{n\rightarrow\infty} I^{\scaleto{DN}{3.5pt}}_n \;,
\end{equation}
\begin{small}
\begin{equation}\label{numDN}
I^{\scaleto{DN}{3.5pt}}_n=\displaystyle{\int\limits_{v(t_0)=0}^{v_{_F}(t)=0}{\prod\limits_{j=1}^{n}\left[\frac{\det{(G_j)}}{(2\pi\varepsilon)^{d}}\right]^{\frac{1}{2}}\prod\limits_{j=1}^{n-1}{\dd v_j}\dd v_{_Vn}\,\,\, e^{-\frac{\varepsilon}{2} \sum\limits_{j=0}^{n}\delta^2 S(q^{\scaleto{DN}{3pt}},v)_j}}}
\end{equation}
\end{small}%
where $v(\tau):=(v_{_V}(\tau),v_{_F}(\tau))$ represents a perturbation around the minimum $q^{\scaleto{DN}{3.5pt}}(\tau)$, and as such it must satisfy null Dirichlet boundary conditions corresponding to the fixed variables. Note that the integration involves only the variable part of the variation at $\tau=t$, namely $v_{_Vn}:=v_{_V}(\tau_n)=v_{_V}(t)$.

Analogously to what was done previously for $\mathcal{K}$ and $\mathcal{N}$, we perform a backward integration of Eq.~(\ref{numDN}) by finding a set of positive definite matrices $F^{\scaleto{DN}{3.5pt}}_j$, that depend on the coefficients of the second variation on $q^{\scaleto{DN}{3.5pt}}(\tau)$, \ie $G$, $A^{\scaleto{DN}{3.5pt}}$ and $V^{\scaleto{DN}{3.5pt}}$, such that (see Appendix~\ref{Ap1intDN})
\begin{equation}\label{res1DN}
{I}^{\scaleto{DN}{3.5pt}}_n=(2\pi)^{\frac{l-d}{2}}\det{\left[\prod\limits_{j=1}^{n} F^{\scaleto{DN}{3.5pt}}_j\right]}^{-\frac{1}{2}} \;. 
\end{equation}

Using the recursion relations for ${I}^{\scaleto{DN}{3.5pt}}_n$ and $F^{\scaleto{DN}{3.5pt}}_j$ we show that the limit in Eq.~(\ref{res1DN}) gives
\begin{equation}\label{pap2DN}
\mathcal{K}_{\scaleto{\mathcal{M}}{3.5pt}}^{sc} = e^{-S(q^{\scaleto{DN}{3pt}})}(2\pi)^{\frac{l-d}{2}}\det\left[{D^{\scaleto{DN}{3.5pt}}}(t_0)\right]^{-\frac{1}{2}}  \;,
\end{equation}
where $D^{\scaleto{DN}{3.5pt}}(\tau)$ solves Eq.~(\ref{papeq}) with $A=A^{\scaleto{DN}{3.5pt}}(\tau)$ and $V=V^{\scaleto{DN}{3.5pt}}(\tau)$. In addition, as a consequence of the backward integration necessary for deriving the matrices $F^{\scaleto{DN}{3.5pt}}_j$, Eq.~(\ref{papeq}) is solved in the backward direction and the boundary conditions (which are reported in Appendix~\ref{Ap1limitDN}) are given in $\tau=t$. Exploiting the link with the Jacobi equation through the non-linear transformation Eq.~(\ref{tran}), we derive the new boundary conditions given in Eq.~(\ref{fin33}) and recover the general expression for marginal transition distributions Eq.~(\ref{fin22}).
Detailed calculations for this last part can be found in Appendix~\ref{JacDN}.

Let us emphasize that when the Lagrangian is quadratic in its variables we have $\rho_{(\scaleto{\mathcal{M}}{3.5pt})}(q_{_{(F)}},t \vert q_0,t_0)=\rho^{\text{sc}}_{(\scaleto{\mathcal{M}}{3.5pt})}(q_{_{(F)}},t \vert q_0,t_0)$, since the second order expansion used in the semi-classical approximation does not neglect any term of higher order.

In conclusion, we have shown how to compute the path integrals appearing in Eq.~(\ref{pathint}) and Eq.~(\ref{pathint2}) in the semi-classical approximation, from the solutions of the Euler-Lagrange equations and of the systems of linear differential equations Eq.~(\ref{fin3}) and Eq.~(\ref{fin33}). 
\vspace{3mm}

\section{Examples}

To show how our approach applies to a number of relevant problems, we present here three illustrative examples. 

To begin, let us summarize briefly the relations between the Langevin and the Fokker-Planck equations, with the associated path integral formulation \cite{GRAHAM, FALKOFF, HAKEN}.

We consider the Langevin equation
\begin{equation}\label{lang}
\dd Q(t)=\mu(Q(t),t)\dd t+\sigma(t)\dd B(t),
\end{equation}
with $Q(t),\,\,\mu(Q(t),t)\in\mathbb{R}^d$, $\sigma(t)\in\mathbb{R}^{d\times l}$ and $B(t)$ an $l-$dimensional standard Wiener process. It is known that the transition probability $\rho(q,t \vert q_0,t_0)$ for the continuous Markovian process $Q(t)$ is the fundamental solution of the Fokker-Planck equation
\begin{equation}\label{fok}
\frac{\partial}{\partial t}\rho=-\frac{\partial}{\partial q_i}\left[\mu_i(q,t)\rho\right]+\frac{1}{2}\Sigma_{ij}(t)\frac{\partial^2}{\partial q_i\partial q_j}\rho
\end{equation}
with $\rho(q,t_0 \vert q_0,t_0)=\delta(q-q_0)$, and where $\Sigma(t)=\sigma(t)\sigma(t)^T\in\mathbb{R}^{d\times d}$ is the diffusion matrix, $\mu(q,t)$ the drift vector, and the Einstein summation convention is adopted for $i,j=1,...,d$. In particular, if $\Sigma$ is constant and $\mu(q)$ is a function of the configuration, $\rho(q,t \vert q_0,t_0)$ has the path integral representation (\ref{pathint}), where the Lagrangian is given by the Onsager-Machlup function \cite{OM}
\begin{equation}\label{OM}
\mathcal{L}(q(\tau),\dot{q}(\tau))=\frac{1}{2}\left[\left(\dot{q}-\mu(q)\right)^T\Sigma^{-1}\left(\dot{q}-\mu(q)\right)\right] \;.
\end{equation}
In the literature, however, an additional factor $+\frac{1}{2}\text{div}\left({\mu(q)}\right)$ usually appears in Eq.~(\ref{OM}). As mentioned before, we observe that this correction is necessary for providing a normalized result when the path integral expression for the transition probability is only defined by $\mathcal{K}$. Our method offers an alternative approach, where we avoid the problem of finding an effective Lagrangian for every application by introducing  explicitly the normalization constant $\mathcal{N}$, see Eq.~(\ref{pathint}).

In the situation where $\Sigma$ is not strictly positive definite, even if the Onsager-Machlup Lagrangian Eq.~(\ref{OM}) is ill-defined, its Hamiltonian form is well defined by $\mathcal{L}(p(\tau))=\frac{1}{2} p^T\Sigma\, p$, where $\dot{q}=\Sigma \,p+\mu(q)$, and $p$ is the conjugate variable of $\dot{q}$ under the Legendre transform. This procedure is justified by taking the limit for a sequence of strictly positive definite matrices converging to $\Sigma$.
At the same time the Hamilton and Jacobi equations for the minima and the fluctuations are also well defined. In particular, the Hamilton equations
\begin{equation}\label{Ham}
\displaystyle{\frac{\dd}{\dd\tau}\begin{pmatrix}
      {q}  \\
     {p}
     \end{pmatrix}
       =J\begin{pmatrix}
       \dd[\mu(q)]^T\,p  \\
     \Sigma \,p+\mu(q)
     \end{pmatrix}}
\end{equation}
are subject to the boundary conditions $q^{\scaleto{D}{3.5pt}}(t_0)=q_0$, $q^{\scaleto{D}{3.5pt}}(t)=q$ for the Dirichlet minimum $q^{\scaleto{D}{3.5pt}}(\tau)$, $q^{\scaleto{N}{3.5pt}}(t_0)=q_0$, $p^{\scaleto{N}{3.5pt}}(t)=0$ for the Neumann minimum $q^{\scaleto{N}{3.5pt}}(\tau)$, and $q^{\scaleto{DN}{3.5pt}}(t_0)=q_0$, $q_{_F}^{\scaleto{DN}{3.5pt}}(t)=q_{_F}$, $p_{_V}^{\scaleto{DN}{3.5pt}}(t)=0$ for the ``mixed'' minimizer $q^{\scaleto{DN}{3.5pt}}(\tau)$. On the other hand, the Jacobi equation in Hamiltonian form presented in Eqs.~(\ref{fin3},\ref{fin33}) is driven by the matrix
\[
E^{\scaleto{i}{5pt}}=\begin{pmatrix}
     \dd^2[\mu(q)]^T\,p\,\,\, \,\,& \dd[\mu(q)]^T\\
     \\
       \dd[\mu(q)]\,\,\,\,\, & \Sigma
       \end{pmatrix}\Bigg\vert_{q^{\scaleto{i}{5pt}},p^{\scaleto{i}{5pt}}}\;, \quad \scaleto{i}{5pt}=\scaleto{D}{3.5pt},\scaleto{N}{3.5pt},\scaleto{DN}{3.5pt}\;.
\]
Here and in Eq.~(\ref{Ham}), $\dd[\mu(q)]$ and $\dd^2[\mu(q)]$ denote respectively the rank-2 and rank-3 tensors of the first and second derivatives in $q$ of the vector field $\mu(q)$.

\subsection{Ornstein-Uhlenbeck process}

As a first application of our method, we consider the d-dimensional Ornstein-Uhlenbeck process \cite{OU,FALKOFF,VATI}, which is described by the Fokker-Planck equation (\ref{fok}) where $\Sigma\in\mathbb{R}^{d\times d}$ is a constant symmetric diffusion matrix and $\mu(q)=-\Theta\, q$, with $\Theta\in\mathbb{R}^{d\times d}$ defines the drift.
It is easy to see that the system is exactly characterised by the same linear Hamilton and Jacobi equation
\begin{equation}\label{OUeq}
\displaystyle{\frac{\dd}{\dd\tau}\begin{pmatrix}
      {w}  \\
     {m}
     \end{pmatrix}
       =\begin{pmatrix}
       -\Theta &\,\,\,\Sigma  \\
      \mathbb{0}  &\,\,\,\Theta^T
     \end{pmatrix}}
     \begin{pmatrix}
      {w}  \\
     {m}
     \end{pmatrix} \;.
\end{equation}
The analytical solution is given by
\begin{align}\label{analy}
w(\tau)&=e^{-\Theta\tau}\left[\int_0^{\tau}{e^{\Theta s}\,\Sigma\, e^{\Theta^T s}C_1}\,\dd s+C_2\right] \;,\\
m(\tau)&=e^{\Theta^T\tau}C_1 \;,
\end{align}
where $C_1$ and $C_2$ are determined from the appropriate boundary conditions. In particular, setting $t_0=0$, we find that $S(q^{\scaleto{N}{3pt}})=0$, and that
\begin{equation}
S(q^{\scaleto{D}{3pt}}) = \frac{1}{2} \hat{q}^T\left[\int_0^{t}{e^{\Theta (s-t)}\,\Sigma\, e^{\Theta^T (s-t)}}\,\dd s\right]^{-1} \hat{q} \;,
\end{equation}
where $\hat{q} = q-e^{-\Theta t}q_0$. The Jacobi fields lead to the factors
\begin{equation}
W^{\scaleto{N}{3.5pt}}(0)=e^{\Theta t},\,\,\,W^{\scaleto{D}{3.5pt}}(0)=\int_0^{t}{e^{\Theta s}\,\Sigma\, e^{\Theta^T (s-t)}}\,\dd s \;.
\end{equation}

Finally, inserting these quantities in Eq.~(\ref{fin2}), we recover the Gaussian transition probability
\begin{small}
\begin{equation}\label{OUGauss}
\rho(q,t \vert q_0,0)=\frac{\exp{[-\frac{1}{2} \hat{q}^T\text{Co}^{-1}(t) \hat{q}]}}{\sqrt{\det[2\pi\,\text{Co}(t)]}} \;,
\end{equation}
\end{small}%
with mean $\text{Av}(q_0,t):=e^{-\Theta t}q_0$, and covariance matrix
\begin{small}
\begin{equation}\label{gaussCov}
\text{Co}(t):=\int_0^{t}{e^{\Theta (s-t)}\,\Sigma\, e^{\Theta^T (s-t)}}\,\dd s \;.
\end{equation}
\end{small}

In addition, notice that the marginal probability density Eq.~(\ref{pathint2}) for the Ornstein-Uhlenbeck process can be derived analytically by means of Eq.~(\ref{fin22}). In fact, simple algebra gives us 
\begin{equation}
S(q^{\scaleto{DN}{3pt}})=\frac{1}{2}(q_{_F}-\text{Av}_{\scaleto{\mathcal{M}}{3.5pt}}(q_0,t))^T\text{Co}_{\scaleto{\mathcal{M}}{3.5pt}}^{-1}(t)(q_{_F}-\text{Av}_{\scaleto{\mathcal{M}}{3.5pt}}(q_0,t)),
\end{equation}
\begin{equation}\label{Wast}
(W^{\scaleto{N}{3.5pt}})^{-1}W^{\scaleto{DN}{3.5pt}}(0)=\begin{pmatrix}
\mathbb{1}_{l\times l} & *\\
\mathbb{0}_{d-l\times l} & \text{Co}_{\scaleto{\mathcal{M}}{3.5pt}}(t)
\end{pmatrix} \;,
\end{equation}
with $\text{Av}_{\scaleto{\mathcal{M}}{3.5pt}}(q_0,t)=(\mathbb{0}_{d-l\times l},\mathbb{1}_{d-l\times d-l})\text{Av}(q_0,t)$ and 
\begin{equation}
\text{Co}_{\scaleto{\mathcal{M}}{3.5pt}}(t)=\begin{pmatrix}
\mathbb{0}_{d-l\times l} & \mathbb{1}_{d-l\times d-l}
\end{pmatrix}\text{Co}(t)\begin{pmatrix}
\mathbb{0}_{l\times d-l}\\
\mathbb{1}_{d-l\times d-l}
\end{pmatrix} \;.
\end{equation}
Since in Eq.~(\ref{Wast}) the entry denoted by ``$*$" is not relevant, Eq.~(\ref{fin22}) automatically implies that
\begin{equation}
\rho_{\scaleto{\mathcal{M}}{3.5pt}}(q_{_F},t \vert q_0,0) = \frac{\exp{[-\frac{1}{2} \hat{q}_{_F}^T \text{Co}_{\scaleto{\mathcal{M}}{3.5pt}}^{-1}(t) \hat{q}_{_F} ]}}{\sqrt{\det[2\pi\,\text{Co}_{\scaleto{\mathcal{M}}{3.5pt}}(t)]}} \;,
\end{equation}
with $\hat{q}_{_F}=q_{_F}-\text{Av}_{\scaleto{\mathcal{M}}{3.5pt}}(q_0,t)$, which is indeed the (Gaussian) marginal of the full transition probability distribution Eq.~(\ref{OUGauss}).

\subsection{Van der Pol oscillator}

\begin{figure}[h!]
  \centering
\includegraphics[width=\columnwidth]{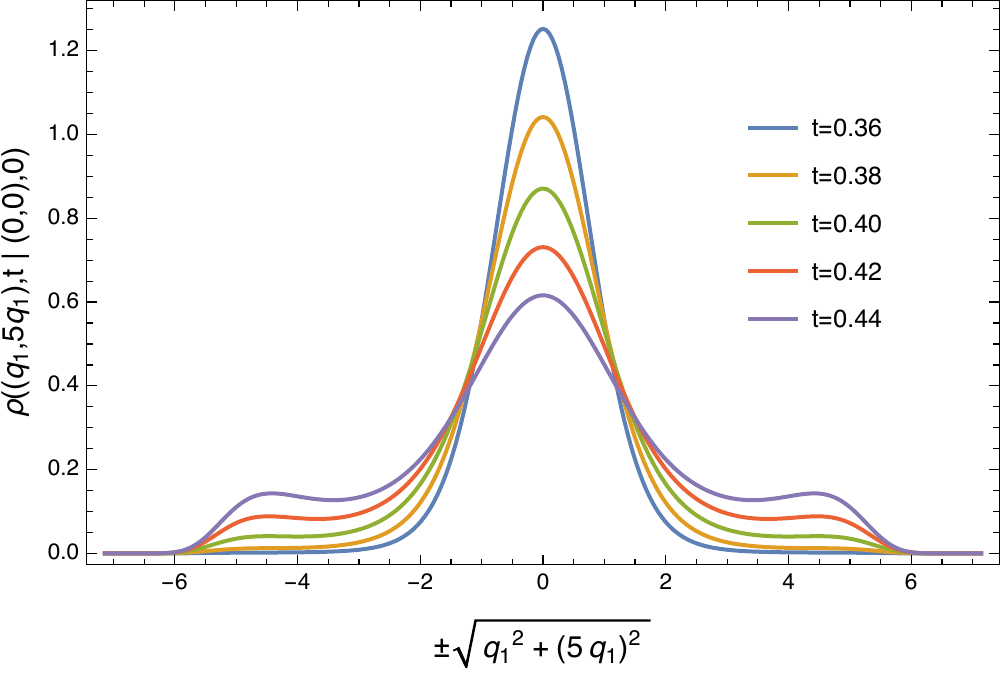}
  \caption{Slices along direction $(q_1,q_2)=(q_1,5 q_1)$ of the transition probability $\rho\left((q_1,q_2),t \vert (0,0),0 \right)$ for the Van der Pol oscillator Eq.~(\ref{vdp}) with $\xi=3$, $\lambda=0.5$. Note that, despite the semi-classical approximation, the resulting probability density is not necessarily Gaussian.}
    \label{fig:vdp}
\end{figure}

As a second application, we consider the Van der Pol oscillator driven by white noise \cite{NAESS}, that is described by the Langevin equation of motion for the coordinate $z$ as
\begin{equation}\label{vdp}
\ddot{z}(t)+2\xi[z(t)^2-1]\dot{z}(t)+z(t)=\sqrt{2 \lambda}f(t) \;,
\end{equation}
where $f(t)$ denotes a standard stationary gaussian white noise, $\lambda>0$ represents the diffusion coefficient and $\xi>0$ the strength of the non-linearity. By defining the terms
\[
\Omega:=\begin{pmatrix}
       0 &\,\,\,-1  \\
      1  &\,\,\,-2\xi
     \end{pmatrix}\,\,\,\,\,\text{and}\,\,\,\,\,\nu(q):=\begin{pmatrix}
       0  \\
      q_1^2q_2
     \end{pmatrix} \;,
\]
it is possible to write the stochastic equation of motion in phase space as a 2-dimensional Langevin equation in the form of Eq.~(\ref{lang}),
with $\sigma=(0,\sqrt{2 \lambda})^T$. The associated Fokker-Planck equation has then coefficients
\[
\Sigma=
\begin{pmatrix}
       0 &\,\,\,0  \\
      0  &\,\,\,2\lambda
     \end{pmatrix},\qquad\text{and}\qquad\mu(q)= 
     -\Omega q-2\xi \nu(q) \;.
\]
Note that, in this example, the corresponding Onsager-Machlup function Eq.~(\ref{OM}) is no longer a quadratic function of $q$ and $\dot{q}$. Therefore, the semi-classical approximation will lead to a result that is \textit{a priori} not exact. From the second order expansion we nevertheless expect the result to be accurate for small values of diffusion $\lambda$ and final time $t$. Applying the method we presented, we obtain that the Hamilton and Jacobi equations for the system are respectively of the form
\begin{equation}\label{Hamvdp}
\displaystyle{\frac{\dd}{\dd\tau}\begin{pmatrix}
      {q}  \\
     {p}
     \end{pmatrix}
       =\begin{pmatrix}
       -\Omega &\,\,\,\Sigma  \\
      \mathbb{0}  &\,\,\,\Omega^T
     \end{pmatrix}
     \begin{pmatrix}
       q  \\
       p
     \end{pmatrix}}-2\xi\psi(q,p) \;,
\end{equation}
\begin{equation}\label{Jacvdp}
\frac{\dd}{\dd\tau}\begin{pmatrix}
      {{W^{\scaleto{i}{5pt}}}}  \\
     {{M^{\scaleto{i}{5pt}}}}
     \end{pmatrix}
       =\begin{pmatrix}
       -\Omega &\,\,\,\Sigma  \\
      \mathbb{0}  &\,\,\,\Omega^T
     \end{pmatrix}
     \begin{pmatrix}
       {{W^{\scaleto{i}{5pt}}}}  \\
     {{M^{\scaleto{i}{5pt}}}}
     \end{pmatrix}
     -2\xi\Psi^{\scaleto{i}{5pt}}\begin{pmatrix}
       {{W^{\scaleto{i}{5pt}}}}  \\
     {{M^{\scaleto{i}{5pt}}}}
     \end{pmatrix} \;,
\end{equation}
with ${\scaleto{i}{5pt}}={\scaleto{D}{3.5pt},\scaleto{N}{3.5pt}}$, $\psi(q,p)=(0,\,q_1^2 q_2,\,-2q_1 q_2 p_2,\,-q_1^2 p_2)^T$, and
\begin{equation*}
     \Psi^{\scaleto{i}{5pt}}=\begin{pmatrix}
       0 &\,\,\,0 &\,\,\,0 &\,\,\,0  \\
       2q_1^{\scaleto{i}{5pt}} q_2^{\scaleto{i}{5pt}} &\,\,\,(q_1^{\scaleto{i}{5pt}})^2 &\,\,\,0 &\,\,\,0  \\
       -2q_2^{\scaleto{i}{5pt}} p_2^{\scaleto{i}{5pt}} &\,\,\,-2q_1^{\scaleto{i}{5pt}} p_2^{\scaleto{i}{5pt}} &\,\,\,0 &\,\,\,-2q_1^{\scaleto{i}{5pt}} q_2^{\scaleto{i}{5pt}}  \\
       -2q_1^{\scaleto{i}{5pt}} p_2^{\scaleto{i}{5pt}} &\,\,\,0 &\,\,\,0 &\,\,\,-(q_1^{\scaleto{i}{5pt}})^2  
     \end{pmatrix} \;.
\end{equation*}
Solving numerically Eqs.~(\ref{Hamvdp},\ref{Jacvdp}), subject to the associated boundary conditions, we are able to obtain through Eq.~(\ref{fin2}) the semi-classical approximation of the (non-Gaussian) transition probability solving the Fokker-Planck equation for the stochastic Van der Pol oscillator, see Fig.\ref{fig:vdp}.

\subsection{String at thermal equilibrium}

As a final example, let us investigate in one dimension the stationary configuration of a charged extensible string at thermal equilibrium, when an external field is applied. We consider the action \cite{Rudi04}
\begin{equation}
    S = \beta \int_0^L \left[ \dfrac{1}{2} \alpha \dot{q}(\tau)^2 - \sigma \phi(q(\tau)) \right] \dd\tau \;,
\end{equation}
where $\beta$ is the inverse temperature, $\alpha$ the elastic constant, $\sigma$ the charge density per unit length, and $\phi(q(\tau))$ the electric potential. Note that here $\tau\in[0,L]$ is a parametrisation of the string, and not a time, so that $\dot{q}(\tau)$ represents the elongation. To be concrete, let us assume a potential of the form $\phi(q)=a q^2 + b q$, which could be the second order approximation of a more general potential. Let us define $\kappa:=\sqrt{2\sigma a / \alpha}$. From the Euler-Lagrange equations with the appropriate boundary conditions we obtain $q^{\scaleto{D}{3.5pt}}(\tau)$ and $q^{\scaleto{N}{3.5pt}}(\tau)$, while from the Jacobi equation we obtain ${W^{\scaleto{D}{3.5pt}}}(\tau)=(\beta\alpha\kappa)^{-1}\sin(\kappa(L-\tau))$ and ${W^{\scaleto{N}{3.5pt}}}(\tau)=\cos(\kappa(L-\tau))$. These results allow us to express the probability that $q(L)=q_L$, given that $q(0)=0$, as
$ p(q_L,L\vert 0,0) = e^{-\frac{1}{2}\frac{(q_L-\text{Av})^2}{\text{Var}}} /\sqrt{2\pi \text{Var}}$, which is a Gaussian probability distribution with mean $\text{Av}:=\frac{b}{2a}\left(\frac{1-\cos(\kappa L)}{\cos(\kappa L)}\right)$ and variance $\text{Var}:=\frac{1}{\beta\alpha \kappa}\tan(\kappa L)$. Interestingly, this example illustrates how $\mathcal{N}\neq 1$ in general, even if $A=0$. Furthermore, let us emphasize that the dynamics of elastic chains is properly described by complex models which allow for the motion in the three dimensional space. In particular, these models typically exploit the concept of framed curve, which takes into account both translational and rotational degrees of freedom.

\section{Conclusions}

In this work we presented a consistent approach to compute transition probabilities in the semi-classical approximation, from a path integral formulation. Our method is based on the generalization of a work by Papadopoulos \cite{PAP1}, which allows us to express the solutions of both conditional and unconditional Gaussian path integrals from the solutions of the Euler-Lagrange equations and a system of linear differential equations. Remarkably, the accuracy of our method is only dependent on the accuracy of the semi-classical approximation. In particular, when the Lagrangian of the system is quadratic in position and velocity there is no approximation, and the results are exact.
As a side note, we discussed what is the effect of choosing different discretisation prescriptions for continuous paths, and mention under which circumstances this can be arbitrary.
To conclude, we applied our method to three examples of general interest. These illustrate how our results can be applied to a variety of problems in physics and mathematics, such as the study of stochastic processes or the analysis of equilibrium configurations of polymers.

\section{Acknowledgments} 
We are grateful to Prof. John Maddocks for the fruitful discussions and insights, as well as to all the LCVMM group of Lausanne for the constant support. GC was supported by the Swiss National Science Foundation through Grant No. 163324. MF acknowledges support by the Swiss National Science Foundation.

\newpage
\appendix
\begin{widetext}

\section{Evaluation of $\mathcal{N}^{\text{sc}}$}

\subsection{Derivation of Eq.~(\ref{res1})}\label{Ap1intN}

In order to derive Eq.~(\ref{res1}), we first express the second variation Eq.~(\ref{secv2}) using the method of finite differences. Recalling that we discretised $\tau\in[t_0,t]$ into $n$ intervals of length $\varepsilon=(t-t_0)/n$, we obtain
\begin{equation}\label{start}
\begin{split}
\delta^2S(q^{\scaleto{N}{3.5pt}},h)\approx\varepsilon\sum\limits_{j=0}^{n}\delta^2S(q^{\scaleto{N}{3.5pt}},h)_j&:=\frac{1}{\varepsilon}\sum\limits_{j=1}^n\left[\Delta {h}_j^TG_j\Delta {h}_j+\varepsilon \Delta {h}_j^T\left({A^{\scaleto{N}{3.5pt}}_j} {h}_j+A^{\scaleto{N}{3.5pt}}_{j-1}{h}_{j-1}\right)+\varepsilon^2 {h}_j^TV^{\scaleto{N}{3.5pt}}_j{h}_j \right]\\
&=\frac{1}{\varepsilon}\sum\limits_{j=1}^n\big{[}{h}_j^T\left(G_j+\varepsilon {A^{\scaleto{N}{3.5pt}}_j}+\varepsilon^2V^{\scaleto{N}{3.5pt}}_j\right){h}_j+{h}_{j-1}^T\left(G_j-\varepsilon A^{\scaleto{N}{3.5pt}}_{j-1}\right){h}_{j-1}\\
&\qquad -{h}_{j}^T\left(G_j-\varepsilon A^{\scaleto{N}{3.5pt}}_{j-1}\right){h}_{j-1}-{h}_{j-1}^T\left(G_j+\varepsilon A^{\scaleto{N}{3.5pt}}_{j}\right){h}_{j}\big{]} \;,
\end{split}
\end{equation}
where we defined $\Delta {h}_j:={h}_j-{h}_{j-1}$ for $j=1,...,n$, and the subscript $j$ indicates that the associated term is evaluated in $\tau_j=t_0+j\varepsilon$ for $j=0,1,...,n$.

Since $h(\tau)$ is a perturbation around the Neumann minimum, then ${h}_0={0}$ and we can rearrange the terms in the sum in order to isolate the slice for $j=n$:
\begin{equation}\label{fin_dif1.2}
\begin{split}
\varepsilon\sum\limits_{j=0}^n\delta^2S(q^{\scaleto{N}{3.5pt}},h)_j&=\frac{1}{\varepsilon}\sum\limits_{j=1}^{n-1}\left[h_j^T\left(G_j+G_{j+1}+\varepsilon^2V^{\scaleto{N}{3.5pt}}_j\right)h_j-h_{j}^T\left(G_j-\varepsilon A^{\scaleto{N}{3.5pt}}_{j-1}\right)h_{j-1}-h_{j-1}^T\left(G_j+\varepsilon A^{\scaleto{N}{3.5pt}}_{j}\right)h_{j}\right]\\
&\quad\quad +\frac{1}{\varepsilon}\left[h_n^T\left(G_n+\varepsilon A^{\scaleto{N}{3.5pt}}_n+\varepsilon^2V^{\scaleto{N}{3.5pt}}_n\right)h_n-h_{n}^T\left(G_n-\varepsilon A^{\scaleto{N}{3.5pt}}_{n-1}\right)h_{n-1}-h_{n-1}^T\left(G_n+\varepsilon A^{\scaleto{N}{3.5pt}}_{n}\right)h_{n}\right] \;.
\end{split}
\end{equation}%
Introducing now the matrices $U^{\scaleto{N}{3.5pt}}_j:=G_j+\frac{\varepsilon}{2}[(A^{\scaleto{N}{3.5pt}}_j)^T-A^{\scaleto{N}{3.5pt}}_{j-1}]$ for $j=1,...,n$, we have that Eq.~(\ref{fin_dif1.2}) can be written as
\begin{small}
\begin{equation}\label{fin_dif2.2}
\begin{split}
\frac{1}{\varepsilon}\sum\limits_{j=1}^{n-1}\left[h_j^T\left(G_j+G_{j+1}+\varepsilon^2V^{\scaleto{N}{3.5pt}}_j\right)h_j-h_j^T U^{\scaleto{N}{3.5pt}}_j h_{j-1}-h_{j-1}^T (U^{\scaleto{N}{3.5pt}}_j)^T h_{j}\right]+\frac{1}{\varepsilon}\left[h_n^T\left(G_n+\varepsilon A^{\scaleto{N}{3.5pt}}_n+\varepsilon^2V^{\scaleto{N}{3.5pt}}_n\right)h_n-h_{n}^T U^{\scaleto{N}{3.5pt}}_n h_{n-1}-h_{n-1}^T (U^{\scaleto{N}{3.5pt}}_n)^Th_{n}\right] \;.
\end{split}
\end{equation}
\end{small}%

At this point, we perform a change of variables. We define the transformation with unit Jacobian ${\phi}_j:=h_j-\beta^{\scaleto{N}{3.5pt}}_j h_{j-1}$ for $j=1,...,n$, where the matrices $\beta^{\scaleto{N}{3.5pt}}_j$ are given recursively by the following construction
\begin{subequations}\label{cons}
\begin{align}
\alpha^{\scaleto{N}{3.5pt}}_{n} &:= G_{n}+\varepsilon\frac{A^{\scaleto{N}{3.5pt}}_n+(A^{\scaleto{N}{3.5pt}}_n)^T}{2}+\varepsilon^2V^{\scaleto{N}{3.5pt}}_{n} \;,&&\\
\alpha^{\scaleto{N}{3.5pt}}_{j} &:= G_{j}+G_{j+1}+\varepsilon^2 V^{\scaleto{N}{3.5pt}}_{j}-(\beta^{\scaleto{N}{3.5pt}}_{j+1})^T\alpha^{\scaleto{N}{3.5pt}}_{j+1}\beta^{\scaleto{N}{3.5pt}}_{j+1} \;&&\text{for }\; j=n-1,...,1 \;, \\
U^{\scaleto{N}{3.5pt}}_j&=\alpha^{\scaleto{N}{3.5pt}}_j\beta^{\scaleto{N}{3.5pt}}_j \;&&\text{for }\; j=1,...,n \;.
\end{align}
\end{subequations}%
These expressions are motivated by the fact that they allow to express Eq.~(\ref{start}) as a sum of quadratic forms, which is desired in view of a Gaussian integration. In fact, ${\phi}_j^T\alpha^{\scaleto{N}{3.5pt}}_j{\phi}_j=h_j^T\alpha^{\scaleto{N}{3.5pt}}_j h_j-h_j^T U^{\scaleto{N}{3.5pt}}_j h_{j-1}-h_{j-1}^T(U^{\scaleto{N}{3.5pt}}_j)^T h_{j}+h_{j-1}^T(\beta^{\scaleto{N}{3.5pt}}_j)^T\alpha^{\scaleto{N}{3.5pt}}_j\beta^{\scaleto{N}{3.5pt}}_j h_{j-1}$, which gives
\begin{equation}\label{quad}
\varepsilon\sum\limits_{j=0}^{n}\delta^2S(q^{\scaleto{N}{3.5pt}},h)_j=\frac{1}{\varepsilon}\sum\limits_{j=1}^{n}{\phi}_j^T\alpha^{\scaleto{N}{3.5pt}}_j{\phi}_j \;.
\end{equation}%
Finally, we define $F^{\scaleto{N}{3.5pt}}_j:=\alpha^{\scaleto{N}{3.5pt}}_jG_j^{-1}$ for $j=1,...,n$, to recover Eq.~(\ref{res1}) by computing the Gaussian integrals as:
\begin{align}
I^{\scaleto{N}{3.5pt}}_n &= \displaystyle{\int\limits_{h(t_0)=0}{\prod\limits_{j=1}^{n}\left[\frac{\det{(G_j)}}{(2\pi\varepsilon)^{d}}\right]^{\frac{1}{2}}{\dd h_j}\,\,\, e^{-\frac{\varepsilon}{2} \sum\limits_{j=0}^{n}\delta^2 S(q^{\scaleto{N}{3.5pt}},h)_j}}} \nonumber\\
&= \displaystyle{\int{\prod\limits_{j=1}^{n}\left[\frac{\det{(G_j)}}{(2\pi\varepsilon)^{d}}\right]^{\frac{1}{2}}{\dd\phi_j}\,\,\, e^{-\frac{1}{2\varepsilon}\sum\limits_{j=1}^{n}{\phi}_j^T\alpha^{\scaleto{N}{3.5pt}}_j{\phi}_j}}} \nonumber\\
&=\det{\left[\prod\limits_{j=1}^{n} F^{\scaleto{N}{3.5pt}}_j\right]}^{-\frac{1}{2}} \;. \label{inttg} 
\end{align}%

\subsection{Derivation of Eq.~(\ref{pap2})}\label{Ap1limitN}

In order to derive Eq.~(\ref{pap2}) we need to compute the limit in Eq.~(\ref{appx2}). To this end, we look for recurrence relations in order to express Eq.~(\ref{inttg}) through a difference equation.
On the basis of the construction given in the previous section, we define 
$D^{\scaleto{N}{3.5pt}}_{n-k}:=\prod\limits_{j=0}^{k}F^{\scaleto{N}{3.5pt}}_{n-j}$ for $k=0,1,...,n-1$, and provide the following iterative method for $D^{\scaleto{N}{3.5pt}}$ and ${\alpha^{\scaleto{N}{3.5pt}}}$.\\
Initial condition: $D^{\scaleto{N}{3.5pt}}_n=\alpha^{\scaleto{N}{3.5pt}}_{n}G_{n}^{-1}$.
Iteration scheme: $D^{\scaleto{N}{3.5pt}}_{n-(k+1)}=D^{\scaleto{N}{3.5pt}}_{n-k}\alpha^{\scaleto{N}{3.5pt}}_{n-(k+1)}G_{n-(k+1)}^{-1}$ for $\,\,k=0,1,...,n-2$.\\
Initial condition: $\alpha^{\scaleto{N}{3.5pt}}_{n}=G_{n}+\varepsilon\frac{A^{\scaleto{N}{3.5pt}}_n+(A^{\scaleto{N}{3.5pt}}_n)^T}{2}+\varepsilon^2 V^{\scaleto{N}{3.5pt}}_{n}$.
Iteration scheme: $\alpha^{\scaleto{N}{3.5pt}}_{n-(k+1)}=G_{n-(k+1)}+G_{n-k}+\varepsilon^2 V^{\scaleto{N}{3.5pt}}_{n-(k+1)}-(\beta^{\scaleto{N}{3.5pt}}_{n-k})^T\alpha^{\scaleto{N}{3.5pt}}_{n-k}\beta^{\scaleto{N}{3.5pt}}_{n-k}$ for $k=0,1,...,n-2$.

Reminding that
$\beta^{\scaleto{N}{3.5pt}}_{n-k}=(\alpha^{\scaleto{N}{3.5pt}}_{n-k})^{-1}U^{\scaleto{N}{3.5pt}}_{n-k}$, and that $U^{\scaleto{N}{3.5pt}}_{n-k}=G_{n-k}+\frac{\varepsilon}{2}[(A^{\scaleto{N}{3.5pt}}_{n-k})^T-A^{\scaleto{N}{3.5pt}}_{n-(k+1)}]$,
it is possible to give the explicit recurrence relation for $\alpha^{\scaleto{N}{3.5pt}}_{n-(k+1)}$, $k=0,1,...,n-2$ as
\begin{equation}\label{rec2}
\begin{split}
\alpha^{\scaleto{N}{3.5pt}}_{n-(k+1)}&=G_{n-(k+1)}+G_{n-k}+\varepsilon^2 V^{\scaleto{N}{3.5pt}}_{n-(k+1)}-G_{n-k}(\alpha^{\scaleto{N}{3.5pt}}_{n-k})^{-1}G_{n-k}\\
&\quad-\varepsilon\left[\frac{A^{\scaleto{N}{3.5pt}}_{n-k}-(A^{\scaleto{N}{3.5pt}}_{n-(k+1)})^T}{2}\right](\alpha^{\scaleto{N}{3.5pt}}_{n-k})^{-1}G_{n-k}-\varepsilon G_{n-k}(\alpha^{\scaleto{N}{3.5pt}}_{n-k})^{-1}\left[\frac{(A^{\scaleto{N}{3.5pt}}_{n-k})^T-A^{\scaleto{N}{3.5pt}}_{n-(k+1)}}{2}\right]\\
&\quad-\varepsilon^2\left[\frac{A^{\scaleto{N}{3.5pt}}_{n-k}-(A^{\scaleto{N}{3.5pt}}_{n-(k+1)})^T}{2}\right](\alpha^{\scaleto{N}{3.5pt}}_{n-k})^{-1}\left[\frac{(A^{\scaleto{N}{3.5pt}}_{n-k})^T-A^{\scaleto{N}{3.5pt}}_{n-(k+1)}}{2}\right].
\end{split}
\end{equation}%
Moreover, the recurrence formula for $D^{\scaleto{N}{3.5pt}}$ provides the additional useful relations
\begin{subequations}\label{Drec}
\begin{align}
\alpha^{\scaleto{N}{3.5pt}}_{n-(k+1)} &= (D^{\scaleto{N}{3.5pt}}_{n-k})^{-1}D^{\scaleto{N}{3.5pt}}_{n-(k+1)}G_{n-(k+1)} \\
(\alpha^{\scaleto{N}{3.5pt}}_{n-k})^{-1} &= G_{n-k}^{-1}(D^{\scaleto{N}{3.5pt}}_{n-k})^{-1}D^{\scaleto{N}{3.5pt}}_{n-(k-1)}.
\end{align}%
\end{subequations}
Finally, substituting Eqs.~(\ref{Drec}) in Eq.~(\ref{rec2}), and multiplying to the left both sides by $D^{\scaleto{N}{3.5pt}}_{n-k}$, we get the full difference equation for the matrix $D^{\scaleto{N}{3.5pt}}$, in terms of $G$, ${A^{\scaleto{N}{3.5pt}}}$ and ${V^{\scaleto{N}{3.5pt}}}$, for $k=1,2,...,n-2$:
\begin{small}
\begin{equation}\label{diff}
\begin{split}
D^{\scaleto{N}{3.5pt}}_{n-(k+1)}G_{n-(k+1)}=&D^{\scaleto{N}{3.5pt}}_{n-k}G_{n-(k+1)}+D^{\scaleto{N}{3.5pt}}_{n-k}G_{n-k}-D^{\scaleto{N}{3.5pt}}_{n-(k-1)}G_{n-k}+\varepsilon^2 D^{\scaleto{N}{3.5pt}}_{n-k}V^{\scaleto{N}{3.5pt}}_{n-(k+1)}\\
&\; -\varepsilon D^{\scaleto{N}{3.5pt}}_{n-k}\left[\frac{A^{\scaleto{N}{3.5pt}}_{n-k}-(A^{\scaleto{N}{3.5pt}}_{n-(k+1)})^T}{2}\right]G_{n-k}^{-1}(D^{\scaleto{N}{3.5pt}}_{n-k})^{-1}D^{\scaleto{N}{3.5pt}}_{n-(k-1)}G_{n-k}-\varepsilon D^{\scaleto{N}{3.5pt}}_{n-(k-1)}\left[\frac{(A^{\scaleto{N}{3.5pt}}_{n-k})^T-A^{\scaleto{N}{3.5pt}}_{n-(k+1)}}{2}\right]\\
&\; -\varepsilon^2 D^{\scaleto{N}{3.5pt}}_{n-k}\left[\frac{A^{\scaleto{N}{3.5pt}}_{n-k}-(A^{\scaleto{N}{3.5pt}}_{n-(k+1)})^T}{2}\right]G_{n-k}^{-1}(D^{\scaleto{N}{3.5pt}}_{n-k})^{-1}D^{\scaleto{N}{3.5pt}}_{n-(k-1)}\left[\frac{(A^{\scaleto{N}{3.5pt}}_{n-k})^T-A^{\scaleto{N}{3.5pt}}_{n-(k+1)}}{2}\right].
\end{split}
\end{equation}
\end{small}%
Our goal is now to take the continuous limit ($n\rightarrow\infty$, $\epsilon\rightarrow 0$) for this expression, in order to obtain a differential equation for the unknown $D^{\scaleto{N}{3.5pt}}$. To this end, remember that \eg $D^{\scaleto{N}{3.5pt}}_{n-(k+1)}$ stands for $D^{\scaleto{N}{3.5pt}}(t-s_{k+1})$ with $s_{k+1}=(k+1)\varepsilon=s_k+\varepsilon$, and that similar expressions hold for all other terms. We can therefore Taylor expand each $D^{\scaleto{N}{3.5pt}}$ around $s_k$ to second order in $\varepsilon$, and each other coefficient to first order. Then, dividing everything by $\varepsilon^2$ we obtain
\begin{small}
\begin{align}
\frac{d}{ds}&\left[\frac{d}{ds}\left[D^{\scaleto{N}{3.5pt}}\small{(t-s)}\right]G(t-s)\right]-D^{\scaleto{N}{3.5pt}}(t-s)\frac{d}{ds}\left[(A^{\scaleto{N}{3.5pt}})^{(s)}(t-s)\right]-D^{\scaleto{N}{3.5pt}}(t-s)\left[V^{\scaleto{N}{3.5pt}}(t-s)+(A^{\scaleto{N}{3.5pt}})^{(a)}(t-s)G^{-1}(t-s)(A^{\scaleto{N}{3.5pt}})^{(a)}(t-s)\right]= \nonumber\\
&=-\frac{d}{ds}\left[D^{\scaleto{N}{3.5pt}}(t-s)\right](A^{\scaleto{N}{3.5pt}})^{(a)}(t-s)+D^{\scaleto{N}{3.5pt}}(t-s)(A^{\scaleto{N}{3.5pt}})^{(a)}(t-s)G^{-1}(t-s)(D^{\scaleto{N}{3.5pt}})^{-1}(t-s)\frac{d}{ds}\left[D^{\scaleto{N}{3.5pt}}(t-s)\right]G(t-s) \;, \label{sys.2}
\end{align}
\end{small}%
subject to the boundary conditions
$D^{\scaleto{N}{3.5pt}}(t-s)\big{|}_{s=0}={\mathbb{1}}$,
$\frac{d}{ds}\left(D^{\scaleto{N}{3.5pt}}\small{(t-s)}\right)\big{|}_{s=0}=(A^{\scaleto{N}{3.5pt}})^{(s)}(t)G(t)^{-1}$.
These are a consequence of the recurrence relations for $D^{\scaleto{N}{3.5pt}}$ and $\alpha^{\scaleto{N}{3.5pt}}$, and they are derived as it follows. For the former we have
\begin{small}
\begin{equation}\label{boun}
\alpha^{\scaleto{N}{3.5pt}}_{n}=G_{n}+\varepsilon\frac{A^{\scaleto{N}{3.5pt}}_n+(A^{\scaleto{N}{3.5pt}}_n)^T}{2} + \varepsilon^2V^{\scaleto{N}{3.5pt}}_{n}\sim G_{n} \,\;\text{as}\;\, \varepsilon\rightarrow 0 \quad\Rightarrow\quad D^{\scaleto{N}{3.5pt}}_{n}  =\alpha^{\scaleto{N}{3.5pt}}_{n}G_{n}^{-1}\sim\mathbb{1} \,\;\text{as}\;\, \varepsilon \rightarrow  0  \;.
\end{equation}%
\end{small}
For the latter, note that
\begin{equation}\label{bounD}
\frac{D^{\scaleto{N}{3.5pt}}_{n-1}-D^{\scaleto{N}{3.5pt}}_{n}}{\varepsilon}=\frac{D^{\scaleto{N}{3.5pt}}_{n}\left(\alpha^{\scaleto{N}{3.5pt}}_{n-1}G_{n-1}^{-1}-{\mathbb{1}}\right)}{\varepsilon}\sim\frac{\left(\alpha^{\scaleto{N}{3.5pt}}_{n-1}G_{n-1}^{-1}-{\mathbb{1}}\right)}{\varepsilon}\,\,\text{as}\,\,\varepsilon\rightarrow 0 \;,
\end{equation}%
and, because $\left[\frac{A^{\scaleto{N}{3.5pt}}_{n}-(A^{\scaleto{N}{3.5pt}}_{n-1})^T+(A^{\scaleto{N}{3.5pt}}_{n})^T-A^{\scaleto{N}{3.5pt}}_{n-1}}{2}\right]G_{n-1}^{-1}\rightarrow 0$ as $\varepsilon\rightarrow 0$, this can also be written as
\begin{equation}\label{Neum}
\begin{split}
\frac{\alpha^{\scaleto{N}{3.5pt}}_{n-1}G_{n-1}^{-1}-{\mathbb{1}}}{\varepsilon}&=\frac{1}{\varepsilon}\big{[}G_{n}G_{n-1}^{-1}+\varepsilon^2 V^{\scaleto{N}{3.5pt}}_{n-1}G_{n-1}^{-1}-G_{n}(\alpha^{\scaleto{N}{3.5pt}}_{n})^{-1}G_{n}G_{n-1}^{-1}\\
&\qquad-\varepsilon\left[\frac{A^{\scaleto{N}{3.5pt}}_{n}-(A^{\scaleto{N}{3.5pt}}_{n-1})^T}{2}\right](\alpha^{\scaleto{N}{3.5pt}}_{n})^{-1}G_{n}G_{n-1}^{-1}-\varepsilon G_{n}(\alpha^{\scaleto{N}{3.5pt}}_{n})^{-1}\left[\frac{(A^{\scaleto{N}{3.5pt}}_{n})^T-A^{\scaleto{N}{3.5pt}}_{n-1}}{2}\right]G_{n-1}^{-1}\\
&\qquad-\varepsilon^2\left[\frac{A^{\scaleto{N}{3.5pt}}_{n}-(A^{\scaleto{N}{3.5pt}}_{n-1})^T}{2}\right](\alpha^{\scaleto{N}{3.5pt}}_{n})^{-1}\left[\frac{(A^{\scaleto{N}{3.5pt}}_{n})^T-A^{\scaleto{N}{3.5pt}}_{n-1}}{2}\right]G_{n-1}^{-1}\big{]}\\
&\sim \frac{1}{\varepsilon}\left[\mathbb{1}-G_{n}(\alpha^{\scaleto{N}{3.5pt}}_{n})^{-1}\right]\,\,\text{as}\,\,\varepsilon\rightarrow 0 \;.
\end{split}
\end{equation}%
Inserting now the definition for $\alpha^{\scaleto{N}{3.5pt}}_{n}$ we have
\begin{equation}\label{bouns}
\frac{1}{\varepsilon}\left[\mathbb{1}-G_{n}(\alpha^{\scaleto{N}{3.5pt}}_{n})^{-1}\right]=\frac{1}{\varepsilon}\left[\mathbb{1}-\left(\mathbb{1}+\varepsilon\frac{A^{\scaleto{N}{3.5pt}}_n+(A^{\scaleto{N}{3.5pt}}_n)^T}{2}G_{n}^{-1}+\varepsilon^2V^{\scaleto{N}{3.5pt}}_{n}G_{n}^{-1}\right)^{-1}\right] \;,
\end{equation}%
which, exploiting the Neumann series $({\mathbb{1}}+\Lambda)^{-1}={\mathbb{1}}-\Lambda+\Lambda^2-\Lambda^3+...$ with $\Lambda=\varepsilon\frac{A^{\scaleto{N}{3.5pt}}_n+(A^{\scaleto{N}{3.5pt}}_n)^T}{2}G_{n}^{-1}+\varepsilon^2V^{\scaleto{N}{3.5pt}}_{n}G_{n}^{-1}$, gives
\begin{equation}\label{finbou}
\frac{D^{\scaleto{N}{3.5pt}}_{n-1}-D^{\scaleto{N}{3.5pt}}_{n}}{\varepsilon}\sim (A^{\scaleto{N}{3.5pt}}_n)^{(s)}G_{n}^{-1}\,\,\text{as}\,\,\varepsilon\rightarrow 0 \;.
\end{equation}%

To summarize, setting $\tau=t-s$ in Eq.~(\ref{sys.2}), this leads to the second order non-linear differential equation Eq.~(\ref{papeq}) subject to the boundary conditions ${D^{\scaleto{N}{3.5pt}}}(t)=\mathbb{1}$ and ${\dot{D}^{\scaleto{N}{3.5pt}}}(t)=-({A^{\scaleto{N}{3.5pt}}})^{\small{(s)}}(t)G(t)^{-1}$.

\section{Different discretisation choices}\label{appB}

According to Eq.~(\ref{disc}), the most general discretisation prescription for Eq.~(\ref{start}) is given as a function of $\gamma\in[0,1]$ as
\begin{equation}\label{startgam}
\begin{split}
\delta^2S(q^{\scaleto{N}{3.5pt}},h)\approx\varepsilon\sum\limits_{j=0}^n\delta^2S(q^{\scaleto{N}{3.5pt}},h)_{\gamma_j}&:=\frac{1}{\varepsilon}\sum\limits_{j=1}^n\left[\Delta {h}_j^TG_j\Delta {h}_j+2\varepsilon \Delta {h}_j^T\left(\gamma{A^{\scaleto{N}{3.5pt}}_j} {h}_j+(1-\gamma)A^{\scaleto{N}{3.5pt}}_{j-1}{h}_{j-1}\right)+\varepsilon^2 {h}_j^TV^{\scaleto{N}{3.5pt}}_j{h}_j \right] \;.\\
\end{split}
\end{equation}%
From this expression, we can repeat all the steps followed in Appendix \ref{Ap1intN} and \ref{Ap1limitN} to see that the difference equation (\ref{diff}) now becomes
\begin{small}
\begin{equation}\label{diff.gam}
\begin{split}
D^{\scaleto{N}{3.5pt}}_{n-(k+1)}G_{n-(k+1)}=&D^{\scaleto{N}{3.5pt}}_{n-k}G_{n-(k+1)}+D^{\scaleto{N}{3.5pt}}_{n-k}G_{n-k}-D^{\scaleto{N}{3.5pt}}_{n-(k-1)}G_{n-k}+\varepsilon^2 D^{\scaleto{N}{3.5pt}}_{n-k}V^{\scaleto{N}{3.5pt}}_{n-(k+1)}+2\varepsilon(2\gamma-1)D^{\scaleto{N}{3.5pt}}_{n-k}A^{\scaleto{N}{3.5pt}}_{n-(k+1)}-\\
&\;-\varepsilon D^{\scaleto{N}{3.5pt}}_{n-k}\left[\gamma{A^{\scaleto{N}{3.5pt}}_{n-k}-(1-\gamma)(A^{\scaleto{N}{3.5pt}}_{n-(k+1)})^T}\right]G_{n-k}^{-1}(D^{\scaleto{N}{3.5pt}}_{n-k})^{-1}D^{\scaleto{N}{3.5pt}}_{n-(k-1)}G_{n-k}-\\
&\;-\varepsilon D^{\scaleto{N}{3.5pt}}_{n-(k-1)}\left[\gamma{(A^{\scaleto{N}{3.5pt}}_{n-k})^T-(1-\gamma)A^{\scaleto{N}{3.5pt}}_{n-(k+1)}}\right]-\\
&\;-\varepsilon^2 D^{\scaleto{N}{3.5pt}}_{n-k}\left[\gamma{A^{\scaleto{N}{3.5pt}}_{n-k}-(1-\gamma)(A^{\scaleto{N}{3.5pt}}_{n-(k+1)})^T}\right]G_{n-k}^{-1}(D^{\scaleto{N}{3.5pt}}_{n-k})^{-1}D^{\scaleto{N}{3.5pt}}_{n-(k-1)}\left[\gamma{(A^{\scaleto{N}{3.5pt}}_{n-k})^T-(1-\gamma)A^{\scaleto{N}{3.5pt}}_{n-(k+1)}}\right].
\end{split}
\end{equation}
\end{small}%

If $A^{\scaleto{N}{3.5pt}}$ is not symmetric, it is possible to derive a differential equation from Eq.~(\ref{diff.gam}) through Taylor expansion only if $\gamma=\frac{1}{2}$, which gives Eq.~(\ref{papeq}).
This is easy to check by performing the calculation. The same is also true for $\mathcal{K}$, for which the expansion is now taken around the isolated minimum $q^{\scaleto{D}{3.5pt}}$.

If $A^{\scaleto{D}{3.5pt}}$ and $A^{\scaleto{N}{3.5pt}}$ are symmetric, and if we adopt the same discretisation prescription for both $\mathcal{K}$ and $\mathcal{N}$, then there is a one parameter family of different equations providing the same normalized result for the transition probability density. Namely
\begin{equation}\label{sys.new.gam}
\frac{\dd}{\dd\tau}\left[\dot{D}{{G}}\right]+2\gamma {{D}}\dot{A}-{{D}}[{{V}}-(1-2\gamma)^2{{A}}{{G}}^{-1}A]=
(1-2\gamma)\dot{D}{{A}}+(1-2\gamma){{D}}{{A}}{{G}}^{-1}D^{-1}\dot{D}G \;,
\end{equation}%
subject to the boundary conditions ${D^{\scaleto{D}{3.5pt}}}(t)=\mathbb{0}$, ${\dot{D}^{\scaleto{D}{3.5pt}}}(t)=-G(t)^{-1}$ and ${D^{\scaleto{N}{3.5pt}}}(t)=\mathbb{1}$, ${\dot{D}^{\scaleto{N}{3.5pt}}}(t)=-2\gamma A^{\scaleto{N}{3.5pt}}(t) G(t)^{-1}$.

\section{The non-linear transformation Eq.~(\ref{tran})}\label{appC}

As mentioned in the main text, the solutions $D$ of Eq.~(\ref{papeq}) and $L$ of Eq.~(\ref{jactr}) are related by the nonlinear transformation Eq.~(\ref{tran}) presented in \cite{LUD}. Here we present in detail how the boundary conditions for $D$ translate into boundary conditions for $L$, in the context of the backward integration procedure.

First, let us consider the case for $q^{\scaleto{D}{3.5pt}}$. Eq.~(\ref{tran}) gives us a mapping between $L^{\scaleto{D}{3.5pt}}$ and $D^{\scaleto{D}{3.5pt}}$, as far as they are invertible. If we assume $L^{\scaleto{D}{3.5pt}}$ and $D^{\scaleto{D}{3.5pt}}$ invertible for all $\tau\neq t$ (no conjugate points) the transformation is valid except for $\tau=t$, where $D^{\scaleto{D}{3.5pt}}(t)=\mathbb{0}$ because of the boundary conditions in the backward direction (see paragraph after Eq.~(\ref{pap2}).

To derive the boundary conditions for $L^{\scaleto{D}{3.5pt}}$ in $\tau=t$ from the boundary conditions for $D^{\scaleto{D}{3.5pt}}$ in $\tau=t$, we consider the following reasoning. For $\tau\neq t$ we can write
$\dot{L^{\scaleto{D}{3.5pt}}}=L^{\scaleto{D}{3.5pt}}(D^{\scaleto{D}{3.5pt}})^{-1}\dot{D^{\scaleto{D}{3.5pt}}}+L^{\scaleto{D}{3.5pt}}(A^{\scaleto{D}{3.5pt}})^{(a)} {G}^{-1}$,
and we know that $D^{\scaleto{D}{3.5pt}}\rightarrow \mathbb{0}$, $\dot{D^{\scaleto{D}{3.5pt}}}\rightarrow -{G}(t)^{-1}$ as $\tau\rightarrow t$, for continuity of $D^{\scaleto{D}{3.5pt}}$ and $\dot{D^{\scaleto{D}{3.5pt}}}$. As a consequence, in order to obtain a finite boundary condition for $\dot{L^{\scaleto{D}{3.5pt}}}$, we necessarily want $L^{\scaleto{D}{3.5pt}}(D^{\scaleto{D}{3.5pt}})^{-1}\rightarrow {X}$ as $\tau\rightarrow t$, where ${X}$ is a finite valued matrix. This implies that $L^{\scaleto{D}{3.5pt}}(t)=\lim_{\tau\rightarrow t}{L^{\scaleto{D}{3.5pt}}}=\mathbb{0}$, for continuity of $L^{\scaleto{D}{3.5pt}}$. Furthermore, having $L^{\scaleto{D}{3.5pt}}$ invertible for $\tau\neq t$ implies $\dot{L^{\scaleto{D}{3.5pt}}}(t)$ is not singular, meaning that the matrix ${X}$ is not singular as well, since $L^{\scaleto{D}{3.5pt}}(t)=\mathbb{0}$. To summarize, we have that
\begin{equation}\label{sysb}
\frac{\dd}{\dd\tau}\left[\dot{L^{\scaleto{D}{3.5pt}}}{G}+L^{\scaleto{D}{3.5pt}}(A^{\scaleto{D}{3.5pt}})^T\right]-\dot{L^{\scaleto{D}{3.5pt}}}A^{\scaleto{D}{3.5pt}}-L^{\scaleto{D}{3.5pt}}V^{\scaleto{D}{3.5pt}}=\mathbb{0}
\end{equation}
is subject to the boundary conditions $L^{\scaleto{D}{3.5pt}}(t)=\mathbb{0}$, $\dot{L^{\scaleto{D}{3.5pt}}}(t)=-{X} {G}(t)^{-1}$.

In the same way, we now consider the case for $q^{\scaleto{N}{3.5pt}}$. Assuming $L^{\scaleto{N}{3.5pt}}$ and $D^{\scaleto{N}{3.5pt}}$ to be non-singular also for $\tau=t$ (note $D^{\scaleto{N}{3.5pt}}(t)=\mathbb{1}$), we have
\begin{align}
L^{\scaleto{N}{3.5pt}}(t)^{-1}\dot{L^{\scaleto{N}{3.5pt}}}(t) &= -(A^{\scaleto{N}{3.5pt}})^{(s)}(t){G}(t)^{-1}+(A^{\scaleto{N}{3.5pt}})^{(a)}(t){G}(t)^{-1} \nonumber\\
&= -(A^{\scaleto{N}{3.5pt}})^T(t){G}(t)^{-1} \;. \label{ttr}
\end{align}
In addition, as we want $L^{\scaleto{N}{3.5pt}}$ to be invertible in $\tau=t$, then ${Y}:=L^{\scaleto{N}{3.5pt}}(t)$ must be a non-singular matrix. To summarize, we have that
\begin{equation}\label{sysbb}
\frac{\dd}{\dd\tau}\left[\dot{L^{\scaleto{N}{3.5pt}}}{G}+L^{\scaleto{N}{3.5pt}}(A^{\scaleto{N}{3.5pt}})^T\right]-\dot{L^{\scaleto{N}{3.5pt}}}A^{\scaleto{N}{3.5pt}}-L^{\scaleto{N}{3.5pt}}V^{\scaleto{N}{3.5pt}}=\mathbb{0}
\end{equation}%
is subject to the boundary conditions $L^{\scaleto{N}{3.5pt}}(t)={Y}$, $\dot{L^{\scaleto{N}{3.5pt}}}(t)=-{Y}(A^{\scaleto{N}{3.5pt}})^T(t){G}(t)^{-1}$.

At this point, we use the observation that $\det{{L}}=c\det{{D}}$ for all $\tau$,
where $c$ is a constant \cite{LUD}. In order to make the transformation unique (up to invertible matrices sharing the same determinant), we impose $c=1$ for both $L^{\scaleto{D}{3.5pt}}$, $D^{\scaleto{D}{3.5pt}}$ and $L^{\scaleto{N}{3.5pt}}$, $D^{\scaleto{N}{3.5pt}}$, which allows us to fix  the matrices $X$ and $Y$. Namely,
\begin{align}
\lim\limits_{\tau\rightarrow t}{\det({L^{\scaleto{D}{3.5pt}}(\tau)(D^{\scaleto{D}{3.5pt}})^{-1}(\tau)})} &= \det({X}) \;,\\
\lim\limits_{\tau\rightarrow t}{\det({L^{\scaleto{N}{3.5pt}}(\tau)(D^{\scaleto{N}{3.5pt}})^{-1}(\tau)})} &= \det({Y})
\end{align}
so that we can set ${X}={Y}={\mathbb{1}}$.

\section{The general formulation for marginals}\label{appD}

\subsection{Derivation of Eq.~(\ref{res1DN})}\label{Ap1intDN}

In order to derive Eq.~(\ref{res1DN}), we first express the second variation of the energy in $q^{\scaleto{DN}{3.5pt}}(\tau)$ using the method of finite differences. Recalling that we discretised $\tau\in[t_0,t]$ into $n$ intervals of length $\varepsilon=(t-t_0)/n$, and that $v(\tau):=(v_{_V}(\tau),v_{_F}(\tau))$ represents a perturbation around the latter minimum, we can start our calculation from expression (\ref{fin_dif2.2}), since the previous steps are identical, obtaining now
\begin{equation}\label{fin_dif2.2DN}
\begin{split}
\varepsilon\sum\limits_{j=0}^n\delta^2S(q^{\scaleto{DN}{3.5pt}},v)_j&=\frac{1}{\varepsilon}\sum\limits_{j=1}^{n-1}\left[v_j^T\left(G_j+G_{j+1}+\varepsilon^2V^{\scaleto{DN}{3.5pt}}_j\right)v_j-v_j^T U^{\scaleto{DN}{3.5pt}}_j v_{j-1}-v_{j-1}^T (U^{\scaleto{DN}{3.5pt}}_j)^T v_{j}\right]\\
&\quad+\frac{1}{\varepsilon}\left[v_n^T\left(G_n+\varepsilon A^{\scaleto{DN}{3.5pt}}_n+\varepsilon^2V^{\scaleto{DN}{3.5pt}}_n\right)v_n-v_{n}^T U^{\scaleto{DN}{3.5pt}}_n v_{n-1}-v_{n-1}^T (U^{\scaleto{DN}{3.5pt}}_n)^Tv_{n}\right] \;,
\end{split}
\end{equation}%
with $U^{\scaleto{DN}{3.5pt}}_j:=G_j+\frac{\varepsilon}{2}[(A^{\scaleto{DN}{3.5pt}}_j)^T-A^{\scaleto{DN}{3.5pt}}_{j-1}]$ for $j=1,...,n$.

To proceed further, let us introduce the following notation. For a given matrix $M\in\mathbb{R}^{d\times d}$, we label the submatrices $\breve{M}\in\mathbb{R}^{l\times l}$, $\hat{M}\in\mathbb{R}^{l\times d-l}$, $\grave{M}\in\mathbb{R}^{d-l\times l}$, $\tilde{M}\in\mathbb{R}^{d-l\times d-l}$ and $\bar{M}\in\mathbb{R}^{l\times d}$, such that $M=\bigl( \begin{smallmatrix} \breve{M} & \hat{M}\\ \grave{M} & \tilde{M}\end{smallmatrix}\bigr)$ and $\bar{M}=(\breve{M}, \hat{M})$.

Since that the variation in $\tau=t$ is given by $v_n=(v_{_Vn},0)$, because of the boundary conditions satisfied by the mixed minimizer $q^{\scaleto{DN}{3.5pt}}(\tau)$, we can write Eq.~(\ref{fin_dif2.2DN}) as
\begin{equation}\label{fin_dif2.2DN2}
\begin{split}
\varepsilon\sum\limits_{j=0}^n\delta^2S(q^{\scaleto{DN}{3.5pt}},v)_j&=\frac{1}{\varepsilon}\sum\limits_{j=1}^{n-1}\left[v_j^T\left(G_j+G_{j+1}+\varepsilon^2V^{\scaleto{DN}{3.5pt}}_j\right)v_j-v_j^T U^{\scaleto{DN}{3.5pt}}_j v_{j-1}-v_{j-1}^T (U^{\scaleto{DN}{3.5pt}}_j)^T v_{j}\right]\\
&\quad+\frac{1}{\varepsilon}\left[v_{_Vn}^T\left(\breve{G}_n+\varepsilon \breve{A}^{\scaleto{DN}{3.5pt}}_n+\varepsilon^2\breve{V}^{\scaleto{DN}{3.5pt}}_n\right)v_{_Vn}-v_{_Vn}^T \bar{U}^{\scaleto{DN}{3.5pt}}_n v_{_V n-1}-v_{_V n-1}^T (\bar{U}^{\scaleto{DN}{3.5pt}}_n)^Tv_{_Vn}\right] \;.
\end{split}
\end{equation}%

At this point, we perform a change of variables. We define the transformation with unit Jacobian ${\phi}_j:=v_j-\beta^{\scaleto{DN}{3.5pt}}_j v_{j-1}\in\mathbb{R}^{d}$ for $j=1,...,n-1$, ${\phi}_n:=v_{_Vn}-\beta^{\scaleto{DN}{3.5pt}}_n v_{n-1}\in\mathbb{R}^{l}$, where the matrices $\beta^{\scaleto{DN}{3.5pt}}_j$ are given recursively by the following construction
\begin{subequations}\label{cons}
\begin{align}
\alpha^{\scaleto{DN}{3.5pt}}_{n} &:= \breve{G}_{n}+\varepsilon\frac{\breve{A}^{\scaleto{DN}{3.5pt}}_n+(\breve{A}^{\scaleto{DN}{3.5pt}}_n)^T}{2}+\varepsilon^2\breve{V}^{\scaleto{DN}{3.5pt}}_{n}\in\mathbb{R}^{l\times l} \;,&&\\
\alpha^{\scaleto{DN}{3.5pt}}_{j} &:= G_{j}+G_{j+1}+\varepsilon^2 V^{\scaleto{DN}{3.5pt}}_{j}-(\beta^{\scaleto{DN}{3.5pt}}_{j+1})^T\alpha^{\scaleto{DN}{3.5pt}}_{j+1}\beta^{\scaleto{DN}{3.5pt}}_{j+1}\in\mathbb{R}^{d\times d} \;&&\text{for }\; j=n-1,...,1 \;, \\
\bar{U}^{\scaleto{DN}{3.5pt}}_n&=\alpha^{\scaleto{DN}{3.5pt}}_n\beta^{\scaleto{DN}{3.5pt}}_n\in\mathbb{R}^{l\times d}\;,\\
U^{\scaleto{DN}{3.5pt}}_j&=\alpha^{\scaleto{DN}{3.5pt}}_j\beta^{\scaleto{DN}{3.5pt}}_j\in\mathbb{R}^{d\times d} \;&&\text{for }\; j=1,...,n-1 \;.
\end{align}
\end{subequations}%
These expressions are motivated by the fact that they allow to express Eq.~(\ref{fin_dif2.2DN2}) as a sum of quadratic forms, which is desired in view of a Gaussian integration, namely
\begin{equation}\label{quad}
\varepsilon\sum\limits_{j=0}^{n}\delta^2S(q^{\scaleto{DN}{3.5pt}},v)_j=\frac{1}{\varepsilon}\sum\limits_{j=1}^{n}{\phi}_j^T\alpha^{\scaleto{DN}{3.5pt}}_j{\phi}_j \;.
\end{equation}%
We then define $F^{\scaleto{DN}{3.5pt}}_j:=\alpha^{\scaleto{DN}{3.5pt}}_jG_j^{-1}$ for $j=1,...,n-1$, and we compute the Gaussian integrals as
\begin{align}
I^{\scaleto{DN}{3.5pt}}_n &=\displaystyle{\int\limits_{v(t_0)=0}^{v_{_F}(t)=0}{\prod\limits_{j=1}^{n}\left[\frac{\det{(G_j)}}{(2\pi\varepsilon)^{d}}\right]^{\frac{1}{2}}\prod\limits_{j=1}^{n-1}{\dd v_j}\dd v_{_Vn}\,\,\, e^{-\frac{\varepsilon}{2} \sum\limits_{j=0}^{n}\delta^2 S(q^{\scaleto{DN}{3pt}},v)_j}}} \nonumber\\
&= \displaystyle{\int{\prod\limits_{j=1}^{n}\left[\frac{\det{(G_j)}}{(2\pi\varepsilon)^{d}}\right]^{\frac{1}{2}}{\dd\phi_j}\,\,\, e^{-\frac{1}{2\varepsilon}\sum\limits_{j=1}^{n}{\phi}_j^T\alpha^{\scaleto{DN}{3pt}}_j{\phi}_j}}} \nonumber\\
&=\left[(2\pi\varepsilon)^{d-l}\det{(\alpha^{\scaleto{DN}{3.5pt}}_n)}\right]^{-\frac{1}{2}}\det{\left[G_n^{-1}\prod\limits_{j=1}^{n-1} F^{\scaleto{DN}{3.5pt}}_j\right]}^{-\frac{1}{2}} \;. \label{inttg} 
\end{align}%
Eq.~(\ref{res1DN}) is finally recovered by making the choice (which substantially simplifies the calculations in the next section)
\begin{equation}
F^{\scaleto{DN}{3.5pt}}_n:=\begin{pmatrix}
\alpha^{\scaleto{DN}{3pt}}_{n} & \hat{G}_n\\ \mathbb{0} & \varepsilon\mathbb{1}_{d-l\times d-l}\end{pmatrix}G_n^{-1} \;.
\end{equation}

\subsection{Derivation of Eq.~(\ref{pap2DN})}\label{Ap1limitDN}

In order to derive Eq.~(\ref{pap2DN}) we need to compute the limit in Eq.~(\ref{appxDN}). We notice that the recurrence relations exploited in Appendix~\ref{Ap1limitN} are the same here, thus leading to the same differential equation (\ref{papeq}). 
Namely, we can define 
$D^{\scaleto{DN}{3.5pt}}_{n-k}:=\prod\limits_{j=0}^{k}F^{\scaleto{DN}{3.5pt}}_{n-j}$ for $k=0,1,...,n-1$, and provide the following iterative method for $D^{\scaleto{DN}{3.5pt}}$ and ${\alpha^{\scaleto{DN}{3.5pt}}}$.\\
Initial condition: $D^{\scaleto{DN}{3.5pt}}_n=F^{\scaleto{DN}{3.5pt}}_{n}$.
Iteration scheme: $D^{\scaleto{DN}{3.5pt}}_{n-(k+1)}=D^{\scaleto{DN}{3.5pt}}_{n-k}\alpha^{\scaleto{DN}{3.5pt}}_{n-(k+1)}G_{n-(k+1)}^{-1}$ for $\,\,k=0,1,...,n-2$.\\
Initial condition: $\alpha^{\scaleto{DN}{3.5pt}}_{n}=\breve{G}_{n}+\varepsilon\frac{\breve{A}^{\scaleto{DN}{3.5pt}}_n+(\breve{A}^{\scaleto{DN}{3.5pt}}_n)^T}{2}+\varepsilon^2\breve{V}^{\scaleto{DN}{3.5pt}}_{n}$.
Iteration scheme: $\alpha^{\scaleto{DN}{3.5pt}}_{n-(k+1)}=G_{n-(k+1)}+G_{n-k}+\varepsilon^2 V^{\scaleto{DN}{3.5pt}}_{n-(k+1)}-(\beta^{\scaleto{DN}{3.5pt}}_{n-k})^T\alpha^{\scaleto{DN}{3.5pt}}_{n-k}\beta^{\scaleto{DN}{3.5pt}}_{n-k}$ for $k=0,1,...,n-2$.
Therefore, the only step left is the computation of the boundary conditions for Eq.~(\ref{sys.2}). 

Using the block matrix inversion formula
\begin{equation}
G_n^{-1}=\begin{pmatrix}
\breve{G}_n & \hat{G}_n\\ \hat{G}_n^T & \tilde{G}_n
\end{pmatrix}^{-1}=\begin{pmatrix}
\breve{G}_n^{-1}+\breve{G}_n^{-1}\hat{G}_n Z_n^{-1}\hat{G}_n^T \breve{G}_n^{-1}&\,\,\,\, -\breve{G}_n^{-1}\hat{G}Z_n^{-1}\\ \\ -Z_n^{-1}\hat{G}_n^T\breve{G}_n^{-1} &\,\,\,\, Z_n^{-1}
\end{pmatrix} \;, \qquad Z=\tilde{G}-\hat{G}^T\breve{G}^{-1}\hat{G} \;,
\end{equation}
and the fact that $\alpha^{\scaleto{DN}{3pt}}_{n}\rightarrow \breve{G}_n$ for $\varepsilon\rightarrow 0$, the following expression is easily obtained 
\begin{equation}\label{boDN}
D^{\scaleto{DN}{3.5pt}}_n=\begin{pmatrix}
\alpha^{\scaleto{DN}{3pt}}_{n} & \hat{G}_n\\ \mathbb{0} & \varepsilon\mathbb{1}_{d-l\times d-l}\end{pmatrix}G_n^{-1}\rightarrow\begin{pmatrix}
\mathbb{1}_{l\times l} & \mathbb{0}_{l\times d-l}\\ \mathbb{0}_{d-l\times l} & \mathbb{0}_{d-l\times d-l}\end{pmatrix}\;, \qquad \varepsilon\rightarrow 0 \;,
\end{equation}
so that for Eq.~(\ref{papeq}) we get $D^{\scaleto{DN}{3pt}}(t)=\bigl( \begin{smallmatrix} \mathbb{1} & \mathbb{0}\\ \mathbb{0} & \mathbb{0}\end{smallmatrix}\bigr)$.
Moreover, the derivative at the boundary is discretised as
\begin{equation}
\frac{D^{\scaleto{DN}{3.5pt}}_{n-1}-D^{\scaleto{DN}{3.5pt}}_{n}}{\varepsilon}=\frac{D^{\scaleto{DN}{3.5pt}}_{n}\left(\alpha^{\scaleto{DN}{3.5pt}}_{n-1}G_{n-1}^{-1}-{\mathbb{1}}\right)}{\varepsilon} \;,
\end{equation}%
where the term
$\varepsilon^{-1}\left(\alpha^{\scaleto{DN}{3.5pt}}_{n-1}G_{n-1}^{-1}-{\mathbb{1}}\right)\sim\varepsilon^{-1}\left(G_{n}-(\beta^{\scaleto{DN}{3.5pt}}_{n})^T\alpha^{\scaleto{DN}{3.5pt}}_{n}\beta^{\scaleto{DN}{3.5pt}}_{n}\right)G_{n-1}^{-1}$
is evaluated by computing\\
\begin{equation}
\begin{split}
(\beta^{\scaleto{DN}{3.5pt}}_{n})^T\alpha^{\scaleto{DN}{3.5pt}}_{n}\beta^{\scaleto{DN}{3.5pt}}_{n}&=(\bar{U}^{\scaleto{DN}{3.5pt}}_n)^T(\alpha^{\scaleto{DN}{3.5pt}}_{n})^{-1}\bar{U}^{\scaleto{DN}{3.5pt}}_n\\
&=\left(\bar{G}_j+\frac{\varepsilon}{2}[(\overline{A^{\scaleto{DN}{3.5pt}}_j)^T}-\bar{A}^{\scaleto{DN}{3.5pt}}_{j-1}]\right)^T(\alpha^{\scaleto{DN}{3.5pt}}_{n})^{-1}\left(\bar{G}_j+\frac{\varepsilon}{2}[(\overline{A^{\scaleto{DN}{3.5pt}}_j)^T}-\bar{A}^{\scaleto{DN}{3.5pt}}_{j-1}]\right)\\
&\sim G^{\alpha}_n+\varepsilon\frac{T_n+T_n^T}{2} \;,
\end{split}
\end{equation}
with
\begin{equation}
G^{\alpha}_n:=\begin{pmatrix}
\breve{G}_n(\alpha^{\scaleto{DN}{3.5pt}}_{n})^{-1}\breve{G}_n &\,\,\, \breve{G}_n(\alpha^{\scaleto{DN}{3.5pt}}_{n})^{-1}\hat{G}_n\\
\\ \hat{G}_n^T(\alpha^{\scaleto{DN}{3.5pt}}_{n})^{-1}\breve{G}_n &\,\,\, \hat{G}_n^T(\alpha^{\scaleto{DN}{3.5pt}}_{n})^{-1}\hat{G}_n\end{pmatrix},\,\,T_n:=
\begin{pmatrix}
\left(\breve{A}^{\scaleto{DN}{3.5pt}}_n-(\breve{A}^{\scaleto{DN}{3.5pt}}_{n-1})^T\right)(\alpha^{\scaleto{DN}{3.5pt}}_{n})^{-1}\breve{G}_n &\,\,\,\,\, \left(\breve{A}^{\scaleto{DN}{3.5pt}}_n-(\breve{A}^{\scaleto{DN}{3.5pt}}_{n-1})^T\right)(\alpha^{\scaleto{DN}{3.5pt}}_{n})^{-1}\hat{G}_n\\ \\ (\grave{A}^{\scaleto{DN}{3.5pt}}_n-\left(\hat{A}^{\scaleto{DN}{3.5pt}}_{n-1})^T\right)(\alpha^{\scaleto{DN}{3.5pt}}_{n})^{-1}\breve{G}_n &\,\,\,\,\, \left(\grave{A}^{\scaleto{DN}{3.5pt}}_n-(\hat{A}^{\scaleto{DN}{3.5pt}}_{n-1})^T\right)(\alpha^{\scaleto{DN}{3.5pt}}_{n})^{-1}\hat{G}_n\end{pmatrix}.
\end{equation}
To conclude, we have
\begin{equation}
\frac{D^{\scaleto{DN}{3.5pt}}_{n-1}-D^{\scaleto{DN}{3.5pt}}_{n}}{\varepsilon}\sim D^{\scaleto{DN}{3.5pt}}_{n}\frac{\left(G_{n}-G^{\alpha}_{n}\right)}{\varepsilon}G_{n-1}^{-1}-D^{\scaleto{DN}{3.5pt}}_{n}\frac{T_n+T_n^T}{2}G_{n-1}^{-1} \;.
\end{equation}
From the definition $T(t):=\lim\limits{_{\varepsilon\rightarrow 0}}{T_n}$, and noting that $\frac{\mathbb{1}-\breve{G}_n(\alpha^{\scaleto{DN}{3pt}}_{n})^{-1}}{\varepsilon}\rightarrow\frac{\breve{A}^{\scaleto{DN}{3pt}}+(\breve{A}^{\scaleto{DN}{3pt}})^T}{2}\breve{G}^{-1}(t)$ for $\varepsilon\rightarrow 0$, we perform the necessary computations that take into account a minus sign when transforming the derivative from Eq.~(\ref{sys.2}) to Eq.~(\ref{papeq}), and finally obtain
\begin{equation}\label{boDDN}
\dot{D}^{\scaleto{DN}{3pt}}(t)=\begin{pmatrix}
-\frac{\breve{A}^{\scaleto{DN}{3pt}}+(\breve{A}^{\scaleto{DN}{3pt}})^T}{2}\left[\breve{G}^{-1}+\breve{G}^{-1}\hat{G} Z^{-1}\hat{G}^T \breve{G}^{-1}\right]&\,\,\,\,\,\,\,\frac{\breve{A}^{\scaleto{DN}{3pt}}+(\breve{A}^{\scaleto{DN}{3pt}})^T}{2}\breve{G}^{-1}\hat{G} Z^{-1}\\
\\Z^{-1}\hat{G}^T\breve{G}^{-1}&\,\,\,\,\,-Z^{-1}\end{pmatrix}(t)+\begin{pmatrix}
\,\overline{\frac{T+T^T}{2}G^{-1}}\,\,\\
\mathbb{0}
\end{pmatrix}(t).
\end{equation}
\vspace{3mm}

\subsection{Boundary conditions for the Jacobi fields}\label{JacDN}

As discussed in Appendix~\ref{appC} for the previous cases, the solutions $D$ of Eq.~(\ref{papeq}) and $L$ of Eq.~(\ref{jactr}) are related by the nonlinear transformation Eq.~(\ref{tran}) presented in \cite{LUD}. Here we explain in detail how the boundary conditions for $D$ translate into boundary conditions for $L$ for the general problem of marginal distributions, and how to recover Eq.~(\ref{fin33}). 

Eq.~(\ref{tran}) gives us a mapping between $L^{\scaleto{DN}{3.5pt}}$ and $D^{\scaleto{DN}{3.5pt}}$, as far as they are invertible. If we assume $L^{\scaleto{DN}{3.5pt}}$ and $D^{\scaleto{DN}{3.5pt}}$ invertible for all $\tau\neq t$ (no conjugate points) the transformation is valid except for $\tau=t$, where $D^{\scaleto{DN}{3.5pt}}$ is singular because of the boundary conditions. Therefore, the first step is to find the Taylor expansion with singular term of $(D^{\scaleto{DN}{3.5pt}})^{-1}(\tau)$ around $\tau=t$, arising from Eqs.~(\ref{boDN}) and (\ref{boDDN}), which leads to
\begin{equation}\label{dm1}
(D^{\scaleto{DN}{3.5pt}})^{-1}(\tau)=P\frac{1}{\tau-t}+R+O(\tau-t),\,\,\,P=\begin{pmatrix}
\breve{P} & \hat{P}\\ \grave{P} & \tilde{P}
\end{pmatrix}=\begin{pmatrix}
\mathbb{0} & \mathbb{0}\\ \mathbb{0} & -Z(t)
\end{pmatrix},\,\,\,R=\begin{pmatrix}
\breve{R} & \hat{R}\\ \grave{R} & \tilde{R}
\end{pmatrix}=\begin{pmatrix}
\mathbb{1} & *\\ * & *
\end{pmatrix},
\end{equation}
where the entries denoted by ``$*$" are unnecessary for the derivation of the results.

Furthermore, Eq.~(\ref{tran}) allows us to write
$\dot{L}^{\scaleto{DN}{3.5pt}}=L^{\scaleto{DN}{3.5pt}}(D^{\scaleto{DN}{3.5pt}})^{-1}\dot{D}^{\scaleto{DN}{3.5pt}}+L^{\scaleto{DN}{3.5pt}}(A^{\scaleto{DN}{3.5pt}})^{(a)} {G}^{-1}$ for $\tau\neq t$. As a consequence, in order to obtain a finite boundary condition for $\dot{L}^{\scaleto{DN}{3.5pt}}$, we necessarily want 
\begin{equation}\label{finX}
L^{\scaleto{DN}{3.5pt}}(D^{\scaleto{DN}{3.5pt}})^{-1}\rightarrow {X}\;\;\text{as}\;\;\tau\rightarrow t \;,
\end{equation}
where ${X}$ is a finite valued matrix.

Since around $\tau=t$ we can also write $L^{\scaleto{DN}{3.5pt}}(\tau)=L^{\scaleto{DN}{3.5pt}}(t)+\dot{L}^{\scaleto{DN}{3.5pt}}(t)(\tau-t)+O(\tau-t)^2$, then for (\ref{finX}) to be true we necessarily want $L^{\scaleto{DN}{3.5pt}}(t)=Y$, with $\hat{Y}=\mathbb{0}$, $\tilde{Y}=\mathbb{0}$, $\breve{X}=\breve{Y}$ and $\grave{X}=\grave{Y}$, so that
\begin{equation}
\lim\limits{_{\tau\rightarrow t}}{L^{\scaleto{DN}{3.5pt}}(D^{\scaleto{DN}{3.5pt}})^{-1}}=\begin{pmatrix}
\breve{Y} & \hat{X}\\ \grave{Y} & \tilde{X}
\end{pmatrix}.
\end{equation}
At this point, we use the observation that $\det{{L}}=c\det{{D}}$ for all $\tau$,
where $c$ is a constant \cite{LUD}. In order to make the transformation unique (up to invertible matrices sharing the same determinant), we impose $c=1$ by choosing $\breve{Y}=\mathbb{1}$, $\tilde{X}=\mathbb{1}$, $\grave{Y}=\mathbb{0}$ and $\hat{X}=\left[(\grave{A}^{\scaleto{DN}{3pt}})^T+\frac{\breve{A}^{\scaleto{DN}{3pt}}-(\breve{A}^{\scaleto{DN}{3pt}})^T}{2}\breve{G}^{-1}\hat{G}\right](t)$. The latter choice will be the key element for deriving the boundary conditions appearing in Eq.~(\ref{fin33}).
We obtain $L^{\scaleto{DN}{3.5pt}}(t)=\bigl( \begin{smallmatrix} \mathbb{1} & \mathbb{0}\\ \mathbb{0} & \mathbb{0}\end{smallmatrix}\bigr)$ and, after some algebra, we arrive at
\begin{equation}
\begin{split}
\dot{L}^{\scaleto{DN}{3.5pt}}(t)&=\begin{pmatrix}
\mathbb{1} &\,\,\,\,\,\,\, (\grave{A}^{\scaleto{DN}{3pt}})^T+\frac{\breve{A}^{\scaleto{DN}{3pt}}-(\breve{A}^{\scaleto{DN}{3pt}})^T}{2}\breve{G}^{-1}\hat{G}\\ \mathbb{0} &\,\,\,\,\,\,\, \mathbb{1}
\end{pmatrix}\dot{D}^{\scaleto{DN}{3.5pt}}(t)+\begin{pmatrix}
\,\overline{(A^{\scaleto{DN}{3.5pt}})^{(a)}G^{-1}}\,\,\\
\mathbb{0}
\end{pmatrix}(t),\\
\\
&=\begin{pmatrix}
-(\breve{A}^{\scaleto{DN}{3pt}})^T\left(\breve{G}^{-1}+\breve{G}^{-1}\hat{G} Z^{-1}\hat{G}^T \breve{G}^{-1}\right)+(\grave{A}^{\scaleto{DN}{3pt}})^T Z^{-1}\hat{G}^T \breve{G}^{-1} &\,\,\,\,\,\,\,\,\,\,(\breve{A}^{\scaleto{DN}{3pt}})^T\breve{G}^{-1}\hat{G} Z^{-1}-(\grave{A}^{\scaleto{DN}{3pt}})^TZ^{-1}\\
Z^{-1}\hat{G}^T \breve{G}^{-1}&\,\,\,\,\,\,\,\,\,\,-Z^{-1}
\end{pmatrix}(t) \;.
\end{split}
\end{equation}

To conclude, reminding that $W^{\scaleto{DN}{3pt}}=(L^{\scaleto{DN}{3pt}})^T$ and that the conjugate variable under the Legendre transform for the Hamiltonian form of the Jacobi equation is given by
$M^{\scaleto{DN}{3pt}}=G\dot{W}^{\scaleto{DN}{3pt}}+A^{\scaleto{DN}{3pt}}W^{\scaleto{DN}{3pt}}$, we recover the boundary conditions for Eq.~(\ref{fin33})
\begin{equation}
W^{\scaleto{DN}{3.5pt}}(t)=\begin{pmatrix}
\mathbb{1}&\mathbb{0}\\
\mathbb{0}&\mathbb{0}
\end{pmatrix},\,\,\,M^{\scaleto{DN}{3.5pt}}(t)=\begin{pmatrix}
\mathbb{0}&\mathbb{0}\\
\mathbb{0}&-\mathbb{1}
\end{pmatrix} \;.
\end{equation}

\end{widetext}
\newpage
\bibliography{paper}

\begin{thebibliography}{33}%
\makeatletter
\providecommand \@ifxundefined [1]{%
 \@ifx{#1\undefined}
}%
\providecommand \@ifnum [1]{%
 \ifnum #1\expandafter \@firstoftwo
 \else \expandafter \@secondoftwo
 \fi
}%
\providecommand \@ifx [1]{%
 \ifx #1\expandafter \@firstoftwo
 \else \expandafter \@secondoftwo
 \fi
}%
\providecommand \natexlab [1]{#1}%
\providecommand \enquote  [1]{``#1''}%
\providecommand \bibnamefont  [1]{#1}%
\providecommand \bibfnamefont [1]{#1}%
\providecommand \citenamefont [1]{#1}%
\providecommand \href@noop [0]{\@secondoftwo}%
\providecommand \href [0]{\begingroup \@sanitize@url \@href}%
\providecommand \@href[1]{\@@startlink{#1}\@@href}%
\providecommand \@@href[1]{\endgroup#1\@@endlink}%
\providecommand \@sanitize@url [0]{\catcode `\\12\catcode `\$12\catcode
  `\&12\catcode `\#12\catcode `\^12\catcode `\_12\catcode `\%12\relax}%
\providecommand \@@startlink[1]{}%
\providecommand \@@endlink[0]{}%
\providecommand \url  [0]{\begingroup\@sanitize@url \@url }%
\providecommand \@url [1]{\endgroup\@href {#1}{\urlprefix }}%
\providecommand \urlprefix  [0]{URL }%
\providecommand \Eprint [0]{\href }%
\providecommand \doibase [0]{http://dx.doi.org/}%
\providecommand \selectlanguage [0]{\@gobble}%
\providecommand \bibinfo  [0]{\@secondoftwo}%
\providecommand \bibfield  [0]{\@secondoftwo}%
\providecommand \translation [1]{[#1]}%
\providecommand \BibitemOpen [0]{}%
\providecommand \bibitemStop [0]{}%
\providecommand \bibitemNoStop [0]{.\EOS\space}%
\providecommand \EOS [0]{\spacefactor3000\relax}%
\providecommand \BibitemShut  [1]{\csname bibitem#1\endcsname}%
\let\auto@bib@innerbib\@empty
%</preamble>
\bibitem [{\citenamefont {Papadopoulos}\ and\ \citenamefont
  {Devreese}(1978)}]{BookPapa}%
  \BibitemOpen
  \bibinfo {editor} {\bibfnamefont {George~J.}\ \bibnamefont {Papadopoulos}}\
  and\ \bibinfo {editor} {\bibfnamefont {J.~T.}\ \bibnamefont {Devreese}},\
  eds.,\ \href {\doibase 10.1007/978-1-4684-9140-1} {\emph {\bibinfo {title}
  {Path Integrals, And Their Applications in Quantum, Statistical and Solid
  State Physics}}}\ (\bibinfo  {publisher} {Springer {US}},\ \bibinfo {year}
  {1978})\BibitemShut {NoStop}%
\bibitem [{\citenamefont {Schulman}(1996)}]{BookSchulman}%
  \BibitemOpen
  \bibfield  {author} {\bibinfo {author} {\bibfnamefont {L.S.}\ \bibnamefont
  {Schulman}},\ }\href@noop {} {\emph {\bibinfo {title} {Techniques and
  Applications of Path Integration}}}\ (\bibinfo  {publisher} {Wiley},\
  \bibinfo {year} {1996})\BibitemShut {NoStop}%
\bibitem [{\citenamefont {Chaichian}\ and\ \citenamefont
  {Demichev}(2001)}]{BookChaichian}%
  \BibitemOpen
  \bibfield  {author} {\bibinfo {author} {\bibfnamefont {M.}~\bibnamefont
  {Chaichian}}\ and\ \bibinfo {author} {\bibfnamefont {A.}~\bibnamefont
  {Demichev}},\ }\href {\doibase 10.1887/0750307137} {\emph {\bibinfo {title}
  {Path Integrals in Physics Volume I Stochastic Processes and Quantum
  Mechanics}}},\ Institute of physics series in mathematical and computational
  physics\ (\bibinfo  {publisher} {Taylor \& Francis},\ \bibinfo {year}
  {2001})\BibitemShut {NoStop}%
\bibitem [{\citenamefont {Zinn-Justin}(2002)}]{BookZinn}%
  \BibitemOpen
  \bibfield  {author} {\bibinfo {author} {\bibfnamefont {Jean}\ \bibnamefont
  {Zinn-Justin}},\ }\href {\doibase 10.1093/acprof:oso/9780198509233.001.0001}
  {\emph {\bibinfo {title} {{Quantum Field Theory and Critical Phenomena; 4th
  ed.}}}},\ Internat. Ser. Mono. Phys.\ (\bibinfo  {publisher} {Clarendon
  Press},\ \bibinfo {address} {Oxford},\ \bibinfo {year} {2002})\BibitemShut
  {NoStop}%
\bibitem [{\citenamefont {Kleinert}(2009)}]{BookKleinert}%
  \BibitemOpen
  \bibfield  {author} {\bibinfo {author} {\bibfnamefont {Hagen}\ \bibnamefont
  {Kleinert}},\ }\href {\doibase 10.1142/7305} {\emph {\bibinfo {title} {{Path
  integrals in quantum mechanics, statistics, polymer physics, and financial
  markets; 5th ed.}}}}\ (\bibinfo  {publisher} {World Scientific},\ \bibinfo
  {address} {Singapore},\ \bibinfo {year} {2009})\BibitemShut {NoStop}%
\bibitem [{\citenamefont {Wiener}(1923)}]{Wiener1}%
  \BibitemOpen
  \bibfield  {author} {\bibinfo {author} {\bibfnamefont {Norbert}\ \bibnamefont
  {Wiener}},\ }\bibfield  {title} {\enquote {\bibinfo {title}
  {Differential-space},}\ }\href {\doibase 10.1002/sapm192321131} {\bibfield
  {journal} {\bibinfo  {journal} {Journal of Mathematics and Physics}\ }\textbf
  {\bibinfo {volume} {2}},\ \bibinfo {pages} {131--174} (\bibinfo {year}
  {1923})}\BibitemShut {NoStop}%
\bibitem [{\citenamefont {Wiener}(1924)}]{Wiener2}%
  \BibitemOpen
  \bibfield  {author} {\bibinfo {author} {\bibfnamefont {Norbert}\ \bibnamefont
  {Wiener}},\ }\bibfield  {title} {\enquote {\bibinfo {title} {{The Average
  value of a Functional*}},}\ }\href {\doibase 10.1112/plms/s2-22.1.454}
  {\bibfield  {journal} {\bibinfo  {journal} {Proceedings of the London
  Mathematical Society}\ }\textbf {\bibinfo {volume} {s2-22}},\ \bibinfo
  {pages} {454--467} (\bibinfo {year} {1924})}\BibitemShut {NoStop}%
\bibitem [{\citenamefont {Feynman}(1948)}]{FEY0}%
  \BibitemOpen
  \bibfield  {author} {\bibinfo {author} {\bibfnamefont {R.~P.}\ \bibnamefont
  {Feynman}},\ }\bibfield  {title} {\enquote {\bibinfo {title} {Space-time
  approach to non-relativistic quantum mechanics},}\ }\href {\doibase
  10.1103/RevModPhys.20.367} {\bibfield  {journal} {\bibinfo  {journal} {Rev.
  Mod. Phys.}\ }\textbf {\bibinfo {volume} {20}},\ \bibinfo {pages} {367--387}
  (\bibinfo {year} {1948})}\BibitemShut {NoStop}%
\bibitem [{\citenamefont {Feynman}\ and\ \citenamefont {Hibbs}(1965)}]{FEY}%
  \BibitemOpen
  \bibfield  {author} {\bibinfo {author} {\bibfnamefont {Richard~Phillips}\
  \bibnamefont {Feynman}}\ and\ \bibinfo {author} {\bibfnamefont
  {Albert~Roach}\ \bibnamefont {Hibbs}},\ }\href@noop {} {\emph {\bibinfo
  {title} {{Quantum mechanics and path integrals}}}},\ International series in
  pure and applied physics\ (\bibinfo  {publisher} {McGraw-Hill},\ \bibinfo
  {address} {New York, NY},\ \bibinfo {year} {1965})\BibitemShut {NoStop}%
\bibitem [{\citenamefont {Edwards}(1965)}]{Edwards1965}%
  \BibitemOpen
  \bibfield  {author} {\bibinfo {author} {\bibfnamefont {S.~F.}\ \bibnamefont
  {Edwards}},\ }\bibfield  {title} {\enquote {\bibinfo {title} {The statistical
  mechanics of polymers with excluded volume},}\ }\href {\doibase
  10.1088/0370-1328/85/4/301} {\bibfield  {journal} {\bibinfo  {journal}
  {Proceedings of the Physical Society}\ }\textbf {\bibinfo {volume} {85}},\
  \bibinfo {pages} {613--624} (\bibinfo {year} {1965})}\BibitemShut {NoStop}%
\bibitem [{\citenamefont {Edwards}(1967)}]{Edwards1967}%
  \BibitemOpen
  \bibfield  {author} {\bibinfo {author} {\bibfnamefont {S~F}\ \bibnamefont
  {Edwards}},\ }\bibfield  {title} {\enquote {\bibinfo {title} {The statistical
  mechanics of polymerized material},}\ }\href {\doibase
  10.1088/0370-1328/92/1/303} {\bibfield  {journal} {\bibinfo  {journal}
  {Proceedings of the Physical Society}\ }\textbf {\bibinfo {volume} {92}},\
  \bibinfo {pages} {9--16} (\bibinfo {year} {1967})}\BibitemShut {NoStop}%
\bibitem [{\citenamefont {Freed}(2007)}]{Freed}%
  \BibitemOpen
  \bibfield  {author} {\bibinfo {author} {\bibfnamefont {Karl~F.}\ \bibnamefont
  {Freed}},\ }\enquote {\bibinfo {title} {Functional integrals and polymer
  statistics},}\ in\ \href {\doibase 10.1002/9780470143728.ch1} {\emph
  {\bibinfo {booktitle} {Advances in Chemical Physics}}}\ (\bibinfo
  {publisher} {John Wiley \& Sons Ltd},\ \bibinfo {year} {2007})\ pp.\ \bibinfo
  {pages} {1--128}\BibitemShut {NoStop}%
\bibitem [{\citenamefont {Papadopoulos}\ and\ \citenamefont
  {Thomchick}(1977)}]{PapaPoly}%
  \BibitemOpen
  \bibfield  {author} {\bibinfo {author} {\bibfnamefont {G~J}\ \bibnamefont
  {Papadopoulos}}\ and\ \bibinfo {author} {\bibfnamefont {J}~\bibnamefont
  {Thomchick}},\ }\bibfield  {title} {\enquote {\bibinfo {title} {On a path
  integral having application in polymer physics},}\ }\href {\doibase
  10.1088/0305-4470/10/7/010} {\bibfield  {journal} {\bibinfo  {journal}
  {Journal of Physics A: Mathematical and General}\ }\textbf {\bibinfo {volume}
  {10}},\ \bibinfo {pages} {1115--1121} (\bibinfo {year} {1977})}\BibitemShut
  {NoStop}%
\bibitem [{\citenamefont {Winkler}\ \emph {et~al.}(1994)\citenamefont
  {Winkler}, \citenamefont {Reineker},\ and\ \citenamefont
  {Harnau}}]{Winkler94}%
  \BibitemOpen
  \bibfield  {author} {\bibinfo {author} {\bibfnamefont {Roland~G.}\
  \bibnamefont {Winkler}}, \bibinfo {author} {\bibfnamefont {Peter}\
  \bibnamefont {Reineker}}, \ and\ \bibinfo {author} {\bibfnamefont {Ludger}\
  \bibnamefont {Harnau}},\ }\bibfield  {title} {\enquote {\bibinfo {title}
  {Models and equilibrium properties of stiff molecular chains},}\ }\href
  {\doibase 10.1063/1.468239} {\bibfield  {journal} {\bibinfo  {journal} {The
  Journal of Chemical Physics}\ }\textbf {\bibinfo {volume} {101}},\ \bibinfo
  {pages} {8119--8129} (\bibinfo {year} {1994})}\BibitemShut {NoStop}%
\bibitem [{\citenamefont {Winkler}\ \emph {et~al.}(1997)\citenamefont
  {Winkler}, \citenamefont {Harnau},\ and\ \citenamefont
  {Reineker}}]{Winkler97}%
  \BibitemOpen
  \bibfield  {author} {\bibinfo {author} {\bibfnamefont {Roland~G.}\
  \bibnamefont {Winkler}}, \bibinfo {author} {\bibfnamefont {Ludger}\
  \bibnamefont {Harnau}}, \ and\ \bibinfo {author} {\bibfnamefont {Peter}\
  \bibnamefont {Reineker}},\ }\bibfield  {title} {\enquote {\bibinfo {title}
  {Distribution functions and dynamical properties of stiff macromolecules},}\
  }\href {\doibase 10.1002/mats.1997.040060603} {\bibfield  {journal} {\bibinfo
   {journal} {Macromolecular Theory and Simulations}\ }\textbf {\bibinfo
  {volume} {6}},\ \bibinfo {pages} {1007--1035} (\bibinfo {year}
  {1997})}\BibitemShut {NoStop}%
\bibitem [{\citenamefont {Vilgis}(2000)}]{Vilgis00}%
  \BibitemOpen
  \bibfield  {author} {\bibinfo {author} {\bibfnamefont {T.A.}\ \bibnamefont
  {Vilgis}},\ }\bibfield  {title} {\enquote {\bibinfo {title} {Polymer theory:
  path integrals and scaling},}\ }\href {\doibase
  https://doi.org/10.1016/S0370-1573(99)00122-2} {\bibfield  {journal}
  {\bibinfo  {journal} {Physics Reports}\ }\textbf {\bibinfo {volume} {336}},\
  \bibinfo {pages} {167 -- 254} (\bibinfo {year} {2000})}\BibitemShut {NoStop}%
\bibitem [{\citenamefont {Cotta-Ramusino}\ and\ \citenamefont
  {Maddocks}(2010)}]{LUD}%
  \BibitemOpen
  \bibfield  {author} {\bibinfo {author} {\bibfnamefont {Ludovica}\
  \bibnamefont {Cotta-Ramusino}}\ and\ \bibinfo {author} {\bibfnamefont
  {John~H.}\ \bibnamefont {Maddocks}},\ }\bibfield  {title} {\enquote {\bibinfo
  {title} {Looping probabilities of elastic chains: A path integral
  approach},}\ }\href {\doibase 10.1103/PhysRevE.82.051924} {\bibfield
  {journal} {\bibinfo  {journal} {Phys. Rev. E}\ }\textbf {\bibinfo {volume}
  {82}},\ \bibinfo {pages} {051924} (\bibinfo {year} {2010})}\BibitemShut
  {NoStop}%
\bibitem [{\citenamefont {Langouche}\ \emph {et~al.}(1982)\citenamefont
  {Langouche}, \citenamefont {Roekaerts},\ and\ \citenamefont
  {Tirapegui}}]{BookLan}%
  \BibitemOpen
  \bibfield  {author} {\bibinfo {author} {\bibfnamefont {F.}~\bibnamefont
  {Langouche}}, \bibinfo {author} {\bibfnamefont {D.}~\bibnamefont
  {Roekaerts}}, \ and\ \bibinfo {author} {\bibfnamefont {E.}~\bibnamefont
  {Tirapegui}},\ }\href@noop {} {\emph {\bibinfo {title} {{Functional
  integration and semiclassical expansions}}}}\ (\bibinfo  {publisher} {Reidel
  Publishing Company},\ \bibinfo {address} {Boston},\ \bibinfo {year}
  {1982})\BibitemShut {NoStop}%
\bibitem [{\citenamefont {Langouche}\ \emph {et~al.}(1981)\citenamefont
  {Langouche}, \citenamefont {Roekaerts},\ and\ \citenamefont
  {Tirapegui}}]{REC}%
  \BibitemOpen
  \bibfield  {author} {\bibinfo {author} {\bibfnamefont {F.}~\bibnamefont
  {Langouche}}, \bibinfo {author} {\bibfnamefont {D.}~\bibnamefont
  {Roekaerts}}, \ and\ \bibinfo {author} {\bibfnamefont {E.}~\bibnamefont
  {Tirapegui}},\ }\bibfield  {title} {\enquote {\bibinfo {title} {Covariant and
  gauge-invariant calculation of higher-order corrections in wkb expansions on
  riemannian manifolds},}\ }\href {\doibase 10.1103/PhysRevD.23.1290}
  {\bibfield  {journal} {\bibinfo  {journal} {Phys. Rev. D}\ }\textbf {\bibinfo
  {volume} {23}},\ \bibinfo {pages} {1290--1304} (\bibinfo {year}
  {1981})}\BibitemShut {NoStop}%
\bibitem [{\citenamefont {DeWitt-Morette}(1976)}]{MOR}%
  \BibitemOpen
  \bibfield  {author} {\bibinfo {author} {\bibfnamefont {Cecile}\ \bibnamefont
  {DeWitt-Morette}},\ }\bibfield  {title} {\enquote {\bibinfo {title} {The
  semiclassical expansion},}\ }\href {\doibase
  https://doi.org/10.1016/0003-4916(76)90041-5} {\bibfield  {journal} {\bibinfo
   {journal} {Annals of Physics}\ }\textbf {\bibinfo {volume} {97}},\ \bibinfo
  {pages} {367 -- 399} (\bibinfo {year} {1976})}\BibitemShut {NoStop}%
\bibitem [{\citenamefont {Papadopoulos}(1975)}]{PAP1}%
  \BibitemOpen
  \bibfield  {author} {\bibinfo {author} {\bibfnamefont {G.~J.}\ \bibnamefont
  {Papadopoulos}},\ }\bibfield  {title} {\enquote {\bibinfo {title} {Gaussian
  path integrals},}\ }\href {\doibase 10.1103/PhysRevD.11.2870} {\bibfield
  {journal} {\bibinfo  {journal} {Phys. Rev. D}\ }\textbf {\bibinfo {volume}
  {11}},\ \bibinfo {pages} {2870--2875} (\bibinfo {year} {1975})}\BibitemShut
  {NoStop}%
\bibitem [{\citenamefont {{Uhlenbeck}}\ and\ \citenamefont
  {{Ornstein}}(1930)}]{OU}%
  \BibitemOpen
  \bibfield  {author} {\bibinfo {author} {\bibfnamefont {G.~E.}\ \bibnamefont
  {{Uhlenbeck}}}\ and\ \bibinfo {author} {\bibfnamefont {L.~S.}\ \bibnamefont
  {{Ornstein}}},\ }\bibfield  {title} {\enquote {\bibinfo {title} {{On the
  Theory of the Brownian Motion}},}\ }\href {\doibase 10.1103/PhysRev.36.823}
  {\bibfield  {journal} {\bibinfo  {journal} {Physical Review}\ }\textbf
  {\bibinfo {volume} {36}},\ \bibinfo {pages} {823--841} (\bibinfo {year}
  {1930})}\BibitemShut {NoStop}%
\bibitem [{\citenamefont {Falkoff}(1958)}]{FALKOFF}%
  \BibitemOpen
  \bibfield  {author} {\bibinfo {author} {\bibfnamefont {David}\ \bibnamefont
  {Falkoff}},\ }\bibfield  {title} {\enquote {\bibinfo {title} {Statistical
  theory of irrversible processes: Part i. intergral over fluctuation path
  formulation},}\ }\href {\doibase
  https://doi.org/10.1016/0003-4916(58)90052-6} {\bibfield  {journal} {\bibinfo
   {journal} {Annals of Physics}\ }\textbf {\bibinfo {volume} {4}},\ \bibinfo
  {pages} {325 -- 346} (\bibinfo {year} {1958})}\BibitemShut {NoStop}%
\bibitem [{\citenamefont {Vatiwutipong}\ and\ \citenamefont
  {Phewchean}(2019)}]{VATI}%
  \BibitemOpen
  \bibfield  {author} {\bibinfo {author} {\bibfnamefont {P.}~\bibnamefont
  {Vatiwutipong}}\ and\ \bibinfo {author} {\bibfnamefont {N.}~\bibnamefont
  {Phewchean}},\ }\bibfield  {title} {\enquote {\bibinfo {title} {Alternative
  way to derive the distribution of the multivariate ornstein--uhlenbeck
  process},}\ }\href {\doibase 10.1186/s13662-019-2214-1} {\bibfield  {journal}
  {\bibinfo  {journal} {Advances in Difference Equations}\ ,\ \bibinfo {pages}
  {276}} (\bibinfo {year} {2019})}\BibitemShut {NoStop}%
\bibitem [{\citenamefont {Naess}\ and\ \citenamefont {Hegstad}(1995)}]{NAESS}%
  \BibitemOpen
  \bibfield  {author} {\bibinfo {author} {\bibfnamefont {A.}~\bibnamefont
  {Naess}}\ and\ \bibinfo {author} {\bibfnamefont {B.~K.}\ \bibnamefont
  {Hegstad}},\ }\bibfield  {title} {\enquote {\bibinfo {title} {Transient and
  stationary response statistics of van der pol oscillators subjected to broad
  band random excitation},}\ }\href {\doibase 10.1007/BF02823198} {\bibfield
  {journal} {\bibinfo  {journal} {Sadhana}\ }\textbf {\bibinfo {volume} {20}},\
  \bibinfo {pages} {389--402} (\bibinfo {year} {1995})}\BibitemShut {NoStop}%
\bibitem [{\citenamefont {Graham}(1977)}]{GRAHAM}%
  \BibitemOpen
  \bibfield  {author} {\bibinfo {author} {\bibfnamefont {Robert}\ \bibnamefont
  {Graham}},\ }\bibfield  {title} {\enquote {\bibinfo {title} {Path integral
  formulation of general diffusion processes},}\ }\href {\doibase
  10.1007/BF01312935} {\bibfield  {journal} {\bibinfo  {journal} {Zeitschrift
  f{\"u}r Physik B Condensed Matter}\ }\textbf {\bibinfo {volume} {26}},\
  \bibinfo {pages} {281--290} (\bibinfo {year} {1977})}\BibitemShut {NoStop}%
\bibitem [{\citenamefont {{Haken}}(1976)}]{HAKEN}%
  \BibitemOpen
  \bibfield  {author} {\bibinfo {author} {\bibfnamefont {H.}~\bibnamefont
  {{Haken}}},\ }\bibfield  {title} {\enquote {\bibinfo {title} {{Generalized
  Onsager-Machlup function and classes of path integral solutions of the
  Fokker-Planck equation and the master equation}},}\ }\href {\doibase
  10.1007/BF01360904} {\bibfield  {journal} {\bibinfo  {journal} {Zeitschrift
  fur Physik B Condensed Matter}\ }\textbf {\bibinfo {volume} {24}},\ \bibinfo
  {pages} {321--326} (\bibinfo {year} {1976})}\BibitemShut {NoStop}%
\bibitem [{\citenamefont {Piterbarg}\ and\ \citenamefont
  {Fatalov}(1995)}]{PIT}%
  \BibitemOpen
  \bibfield  {author} {\bibinfo {author} {\bibfnamefont {V~I}\ \bibnamefont
  {Piterbarg}}\ and\ \bibinfo {author} {\bibfnamefont {V~R}\ \bibnamefont
  {Fatalov}},\ }\bibfield  {title} {\enquote {\bibinfo {title} {The laplace
  method for probability measures in banach spaces},}\ }\href {\doibase
  10.1070/rm1995v050n06abeh002635} {\bibfield  {journal} {\bibinfo  {journal}
  {Russian Mathematical Surveys}\ }\textbf {\bibinfo {volume} {50}},\ \bibinfo
  {pages} {1151--1239} (\bibinfo {year} {1995})}\BibitemShut {NoStop}%
\bibitem [{\citenamefont {Gelfand}\ and\ \citenamefont {Fomin}(2012)}]{FOM}%
  \BibitemOpen
  \bibfield  {author} {\bibinfo {author} {\bibfnamefont {I.~M.}\ \bibnamefont
  {Gelfand}}\ and\ \bibinfo {author} {\bibfnamefont {S.~V.}\ \bibnamefont
  {Fomin}},\ }\bibfield  {title} {\enquote {\bibinfo {title} {Calculus of
  variations},}\ }\href@noop {} {\  (\bibinfo {year} {2012})}\BibitemShut
  {NoStop}%
\bibitem [{\citenamefont {Langouche}\ \emph
  {et~al.}(1979{\natexlab{a}})\citenamefont {Langouche}, \citenamefont
  {Roekaerts},\ and\ \citenamefont {Tirapegui}}]{LANG1}%
  \BibitemOpen
  \bibfield  {author} {\bibinfo {author} {\bibfnamefont {F.}~\bibnamefont
  {Langouche}}, \bibinfo {author} {\bibfnamefont {D.}~\bibnamefont
  {Roekaerts}}, \ and\ \bibinfo {author} {\bibfnamefont {E.}~\bibnamefont
  {Tirapegui}},\ }\bibfield  {title} {\enquote {\bibinfo {title} {Functional
  integral methods for stochastic fields},}\ }\href {\doibase
  https://doi.org/10.1016/0378-4371(79)90054-2} {\bibfield  {journal} {\bibinfo
   {journal} {Physica A: Statistical Mechanics and its Applications}\ }\textbf
  {\bibinfo {volume} {95}},\ \bibinfo {pages} {252 -- 274} (\bibinfo {year}
  {1979}{\natexlab{a}})}\BibitemShut {NoStop}%
\bibitem [{\citenamefont {Langouche}\ \emph
  {et~al.}(1979{\natexlab{b}})\citenamefont {Langouche}, \citenamefont
  {Roekaerts},\ and\ \citenamefont {Tirapegui}}]{LANG2}%
  \BibitemOpen
  \bibfield  {author} {\bibinfo {author} {\bibfnamefont {F.}~\bibnamefont
  {Langouche}}, \bibinfo {author} {\bibfnamefont {D.}~\bibnamefont
  {Roekaerts}}, \ and\ \bibinfo {author} {\bibfnamefont {E.}~\bibnamefont
  {Tirapegui}},\ }\bibfield  {title} {\enquote {\bibinfo {title} {Functional
  integrals and the fokker-planck equation},}\ }\href {\doibase
  10.1007/BF02739307} {\bibfield  {journal} {\bibinfo  {journal} {Il Nuovo
  Cimento B (1971-1996)}\ }\textbf {\bibinfo {volume} {53}},\ \bibinfo {pages}
  {135--159} (\bibinfo {year} {1979}{\natexlab{b}})}\BibitemShut {NoStop}%
\bibitem [{\citenamefont {Onsager}\ and\ \citenamefont {Machlup}(1953)}]{OM}%
  \BibitemOpen
  \bibfield  {author} {\bibinfo {author} {\bibfnamefont {L.}~\bibnamefont
  {Onsager}}\ and\ \bibinfo {author} {\bibfnamefont {S.}~\bibnamefont
  {Machlup}},\ }\bibfield  {title} {\enquote {\bibinfo {title} {Fluctuations
  and irreversible processes},}\ }\href {\doibase 10.1103/PhysRev.91.1505}
  {\bibfield  {journal} {\bibinfo  {journal} {Phys. Rev.}\ }\textbf {\bibinfo
  {volume} {91}},\ \bibinfo {pages} {1505--1512} (\bibinfo {year}
  {1953})}\BibitemShut {NoStop}%
\bibitem [{\citenamefont {Podgornik}(2004)}]{Rudi04}%
  \BibitemOpen
  \bibfield  {author} {\bibinfo {author} {\bibfnamefont {Rudi}\ \bibnamefont
  {Podgornik}},\ }\bibfield  {title} {\enquote {\bibinfo {title}
  {Polyelectrolyte-mediated bridging interactions},}\ }\href {\doibase
  10.1002/polb.20205} {\bibfield  {journal} {\bibinfo  {journal} {Journal of
  Polymer Science Part B: Polymer Physics}\ }\textbf {\bibinfo {volume} {42}},\
  \bibinfo {pages} {3539--3556} (\bibinfo {year} {2004})}\BibitemShut {NoStop}%
\end{thebibliography}%

\end{document}